\documentclass[journal,10pt]{IEEEtran}
\pdfoutput=1
\IEEEoverridecommandlockouts
\usepackage{amsbsy}
\usepackage{amsthm}
\usepackage{amssymb}
\usepackage{amsmath}
\usepackage{multirow}
\usepackage{epsfig}
\usepackage{subfigure}
\usepackage{graphics}
\usepackage{multicol}

\usepackage{epstopdf}
\usepackage{balance}
\usepackage{xcolor}
\usepackage{cite}
\usepackage{color}
\usepackage{algorithm}
\usepackage{algorithmic}
\usepackage{textcomp}
\usepackage{subfigure}
\usepackage{subfloat}
\usepackage{booktabs}
\usepackage{bm}
\usepackage{filecontents}

\graphicspath{ {figures/}}

\begin{document}

\title{Sensing as a Service in 6G Perceptive Networks: A Unified Framework for ISAC Resource Allocation}

\author{Fuwang Dong,~\IEEEmembership{Member,~IEEE}, Fan Liu,~\IEEEmembership{Member,~IEEE}, Yuanhao Cui,~\IEEEmembership{Member,~IEEE}, Wei Wang,~\IEEEmembership{Senior~Member,~IEEE}, Kaifeng Han,~\IEEEmembership{Member,~IEEE}, and Zhiqin Wang

\thanks{This work was supported in part by the National Key R\&D Program of China (No. 2021YFB2900200), in part by the National Natural Science Foundation of China under Grant 62101234 and Grant U20B2039, and in part by the Young Elite Scientist Sponsorship Program by the China Association for Science and Technology (CAST) under Grant No. YESS20210055. (\textit{Corresponding author: Fan Liu, Zhiqin Wang.})}
\thanks{F. Dong and F. Liu are with the Department of Electronic and Electrical Engineering, Southern University of Science and Technology, Shenzhen 518055, China (email: dongfw@sustech.edu.cn; liuf6@sustech.edu.cn)}
\thanks{Y. Cui is with the School of Information and Communication Engineering, Beijing University of Posts and Telecommunications, Beijing 100876, China (email: cuiyuanhao@bupt.edu.cn)}
\thanks{W. Wang is with the College of Intelligent System Science and Engineering, Harbin Engineering University, Harbin 150001, China (email: wangwei407@hrbeu.edu.cn)}
\thanks{K. Han and Z. Wang are with the China Academy of Information and Communications Technology, Beijing 100191, China (email: \{hankaifeng, zhiqing.wang\}@caict.ac.cn).}
}

\maketitle

\begin{abstract}
In the upcoming next-generation (5G-Advanced and 6G) wireless networks, sensing as a service will play a more important role than ever before. Recently, the concept of perceptive network is proposed as a paradigm shift that provides sensing and communication (S\&C) services simultaneously. This type of technology is typically referred to as Integrated Sensing and Communications (ISAC). In this paper, we propose the concept of sensing quality of service (QoS) in terms of diverse applications. Specifically, the probability of detection, the Cr\'amer-Rao bound (CRB) for parameter estimation and the posterior CRB for moving target indication are employed to measure the sensing QoS for detection, localization, and tracking, respectively. Then, we establish a unified framework for ISAC resource allocation, where the \textit{fairness} and the \textit{comprehensiveness} optimization criteria are considered for the aforementioned sensing services. The proposed schemes can flexibly allocate the limited power and bandwidth resources according to both S\&C QoSs. Finally, we study the performance trade-off between S\&C services in different resource allocation schemes by numerical simulations.      
\end{abstract}

\IEEEpeerreviewmaketitle

\section{Introduction}\label{Introduction}

\IEEEPARstart{C}{ommunication} quality of service (QoS) is the main indicator concerned in the wireless cellular networks during the past decades. In addition to that, it is well-recognized that sensing service will play a more important role than ever before in the upcoming 5G-Advanced and 6G networks, especially for environment-aware applications \cite{Overview2020}. While current cellular networks may provide sensing functions to a certain extent, e.g., localization, there are two main drawbacks prohibiting future applications. On the one hand, the resolution and localization accuracy provided by the conventional cellular technology are unable to meet the demand for high-precision sensing. On the other hand, the state-of-the-art cellular localization technologies are mostly implemented in a device-based manner, where a signaling device is attached to the object to be located, and is thereby challenging to generalize to broader scenarios which require to sense device-free objects. Fortunately, the recent proposed perceptive network concept, in conjunction with the \textit{Integrated Sensing and Communication (ISAC)} technology, is expected to provide both robust sensing and wireless connectivity \cite{ZAdrew2021}. In such a network, the cellular system is equipped with the networked sensing capability, providing various sensing services to a vast number of users to enable numerous emerging applications, including connected vehicles, drone monitoring, and smart manufacturing. 
Below we briefly overview the ISAC technology adopted in the perceptive network.

\subsection{Integrated Sensing and Communication}\label{IntroISAC}
Motivated by the need of extra frequency spectrum resources for the unprecedented proliferation of new wireless services, ISAC was initially proposed to release the frequency bands previously restricted to radar systems for shared use of communications \cite{ZL2019,CC2020}. The relevant research works span from radar and communication spectral coexistence, radar-communication cooperation, to dual-functional radar-communication (DFRC) design \cite{DW2017,LHzo2020,ComprehensiveS2021}. Recently, the concept and scope of ISAC have been formally given in \cite{liu2021integrated,CuiNet}, which attracts extensive research attention from both industry \cite{Huawei2021} and academia \cite{ZJ2021}. In contrast to the dedicated sensing or communication functionality, the ISAC design methodology exhibits two types of gains. First, the shared use of limited resources, namely, spectrum, energy and hardware platforms, results in improved efficiency for both sensing and communication (S\&C), and hence provides \textit{integration gain}. Second, mutual assistance between S\&C may further boost the dual performance, offering \textit{coordination gain} \cite{liu2021integrated}. Due to the numerous advantages offered by ISAC, it is envisioned to be a key enabler for many future applications including intelligent connected vehicles, Internet of Things, and smart homes and cities \cite{Automous2020,IoT2021}.
 
To evaluate the performance of a perceptive network, QoS metrics are needed for both S\&C services. While the communication QoS has been well studied from both efficiency (spectral and energy efficiency) and reliability (bit and symbol error rates), to the best of our knowledge, the sensing QoS still remains widely unexplored. In what follows, we first define sensing QoS for diverse applications.

\subsection{The Definition of Sensing QoS}\label{DefinitionQoS}
In general, sensing tasks can be roughly classified into four categories, i.e., detection, localization and tracking, imaging, and recognition \cite{liu2021integrated}. In this paper, we focus on the fundamental sensing QoS including the capability to detect, localize and track objects, and designate the QoS definition of the rest of sensing tasks as our future research.

\textit{(1) Detection QoS:} Target detection refers to making binary or multiple decisions to identify the status of a target, e.g., present or absent. The common metrics include the probability of detection $P_\text{D}$, i.e., the probability that a target is declared when a target is in fact present, and the probability of false alarm $P_\text{FA}$, i.e., the probability that a target is declared but in fact absent. In radar application, it usually requires that the $P_\text{FA}$ has to be maintained below a pre-assigned threshold while maximizing the $P_\text{D}$, namely, the Neyman-Pearson criterion \cite{BookDetectionTheory}. A detailed description can be found in Section \ref{TarDetection}.

\textit{(2) Localization QoS:} The localization of the static objects can be interpreted as parameter estimation problems for the time delay and angle of arrival (AoA). A straightforward metric to measure the localization performance is the mean squared error (MSE) between the true parameters and the estimated ones. However, the MSE is normally difficult to characterize. Alternatively, the Cr\'amer-Rao bound (CRB) for target estimation, which is known as a lower bound on the variance of an unbiased estimator, can be employed to measure the localization QoS \cite{BookEstimationTheory}. This will be detailed in Section \ref{TarLocalization}.    

\textit{(3) Tracking QoS:} Target tracking refers to tracking the state variation (range, angle, velocity, etc.) of a moving target, e.g., a vehicle or a drone. Tracking tasks typically emerge in high-mobility ISAC scenarios, such as vehicle-to-everything (V2X) networks. In contrast to the conventional CRB that relies on the measured data only, a posterior Cr\'amer-Rao bound (PCRB) is introduced to measure the tracking QoS by considering the Fisher information provided by both the measured data as well as prior state models \cite{PCRB1998}. A detailed description can be found in Section \ref{TarTracking}.
 
Similar to the conventional communication-only scenarios, users in a perceptive network may also require different sensing QoS. For instance, important/sensitive targets (such as pedestrians, fast-moving vehicles, etc.) generally require high sensing QoS to prevent potential traffic accidents and to safeguard human lives. On the contrary, low sensing QoS could be tolerable to the static and inanimate objects. To this end, the base station (BS) can allocate the available resources to the users who have different S\&C needs, to improve the flexibility and capacity of the perceptive networks.
     
\subsection{Resource Allocation}\label{IntroRA}

Intuitively, the more transmit resource is utilized to serve multiple users, the better S\&C QoS will be obtained. Unfortunately, resources (e.g., transmit power, bandwidth, etc.) are always limited for practical applications, resulting in a performance trade-off among the users. To that end, efficient resource allocation (RA) schemes are necessary to achieve optimal QoS performance. 

On the one hand, the communication achievable rate depends on the power and bandwidth resources based on the Shannon's theorem. Hence, the bandwidth allocation and power control have been widely investigated to maximize the network's capacity \cite{KDcom2003,Propor2005,LTERA2013,GRA2017,NOMARA2018}. In particular, for the orthogonal frequency division multiplexing (OFDM) systems, the water-filling algorithm over inverse of the channel spectrum can be used to optimally allocate power and rate among subcarriers \cite{KDcom2003}. In \cite{Propor2005}, the authors further proposed a joint subchannel assignment and power allocation algorithm under the proportional fairness constraints. Moreover, the RA problems have also been extensively studied in various scenarios, including in the device-to-device communications in LTE-Advanced networks \cite{LTERA2013}, the slicing network based on 5G \cite{GRA2017}, and the non-orthogonal multiple access (NOMA) scheme \cite{NOMARA2018}.

On the other hand, there is relatively less literature on the RA schemes for sensing compared to communications. In \cite{Yan2015}, the simultaneous multibeam and power allocation is developed to improve the worst-case tracking performance, where the number of beams is assumed less than the number of radiating elements. Additionally, the dwell time which determines the velocity estimation accuracy, is regarded as a limited resource to be allocated in phased array radar network \cite{Yan2017cooperative}. In \cite{ZHW2020}, the authors propose a power and bandwidth allocation method to minimize the PCRB of multiple targets in the tracking strategy.  

There have been a number of works investigating the RA problems in ISAC systems. In \cite{SPL2017}, a DFRC transmitter that supports both communication and radar receivers was considered, where the power budget is allocated to radar waveform and information signal such that the probability of detection is maximized under the constraint of information rate requirement. In \cite{LY2017}, a power allocation scheme was proposed in the OFDM ISAC system, where the conditional mutual information (MI) between the random target impulse response and the received signal is adopted as the sensing performance metric. Although the conditional MI is a ``communication-friendly'' metric that has a similar expression to the communication information rate, and is applicable to all kinds of sensing tasks, the vague physical definition and weak correlation with traditional radar metrics limits its practical application. Moreover, the above-mentioned works mainly focused on power allocation for specific ISAC scenarios, without addressing the more general RA problem at a network level. The joint RA scheme for different kinds of system resources (e.g., power and bandwidth) for ISAC systems and networks still remains widely unexplored.

 \begin{figure*}[!t]
	\centering
	\includegraphics[width=6in]{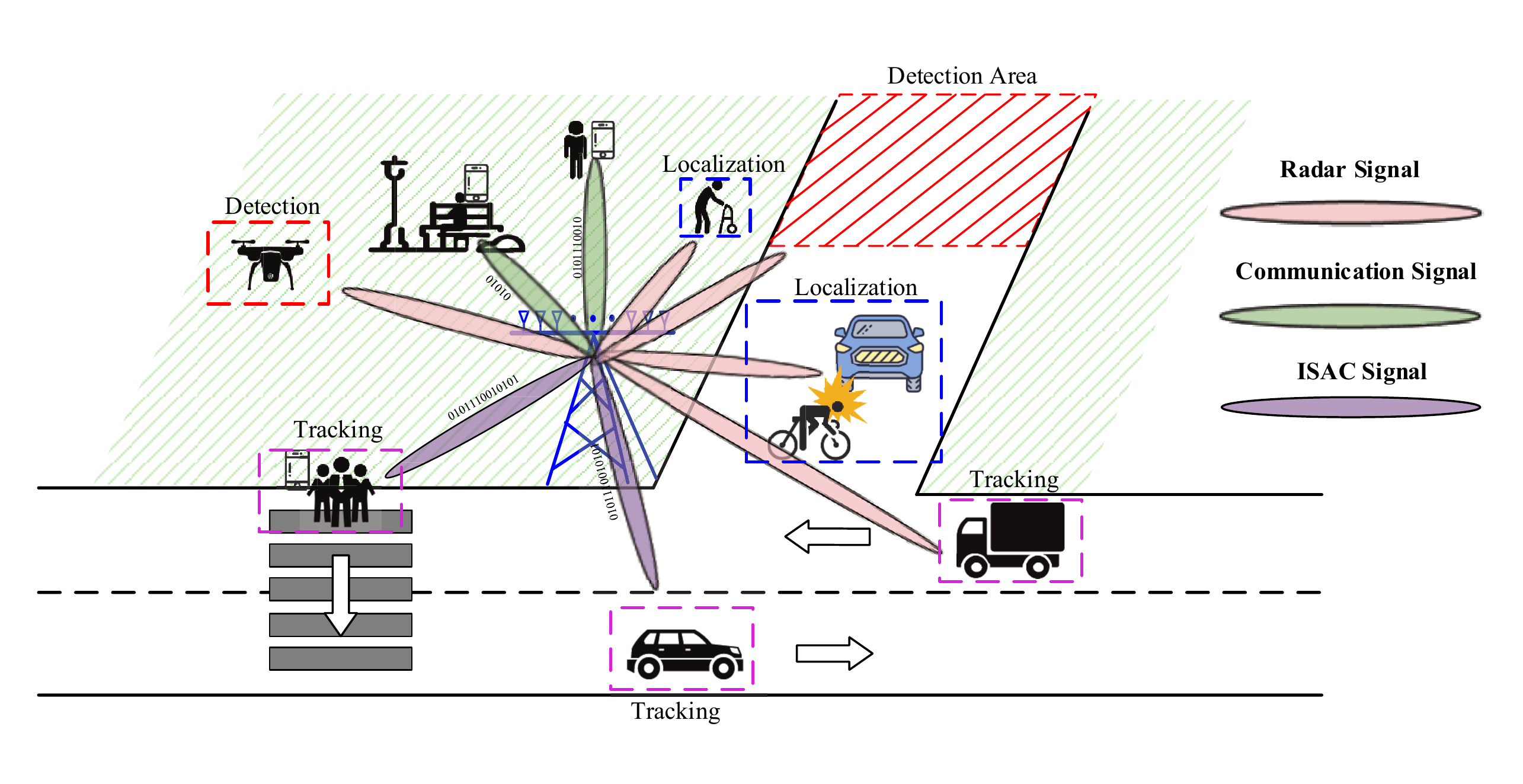}
	\caption{The application scenarios of a single ISAC BS, providing communication and device-free sensing services simultaneously.}
	\label{system_diagram}
\end{figure*}

\subsection{Our Contributions}

Different from the conventional RA schemes that only take the performance trade-off among communication users or radar targets into account, the trade-off between S\&C services should also be considered in a perceptive network, in order to provide S\&C services tailored for user's specific QoS demands. Furthermore, the importance and motivation of the same resource can be quite different between S$\&$C services. This leads to unique challenges and opportunities in the design of ISAC enabled next-generation wireless networks. To fill this research gap, we consider a unified framework for joint power and bandwidth allocation in a single-cell ISAC system, where a multi-input multi-output (MIMO) BS provides communication and device-free sensing services simultaneously. The cooperative sensing for multi-cell ISAC systems, where the clock synchronization and data fusion techniques should be further considered \cite{XYF2018,XYF2022}, is left for our future research. The main contributions of this paper are summarized as follows.

\begin{itemize}
	\item First, we establish a general framework for ISAC RA designs in the upcoming 6G perceptive networks, where the trade-offs between the communication service and various sensing services (e.g., detection, localization, and tracking) are discussed in detail. In the proposed scheme, the BS can provide communication and device-free sensing services tailored for user-specified QoS demands. Additionally, we provide novel insights into a promising use case in perceptive networks. 
 	
	\item Second, we explore the definitions of sensing QoS for different tasks to efficiently evaluate the sensing performance. Specifically, the $P_\text{D}$ and CRB are used to measure the sensing QoS for target detection and localization. As for tracking, we establish the state and measurement models to derive the expression of PCRB as the sensing QoS, where an extended Kalman filtering scheme is adopted to construct a closed tracking loop. 
	  
	\item Third, we propose the effective approaches to solve the RA problems where both the \textit{fairness} and \textit{comprehensiveness} criteria are considered in each sensing task. The original problems are transformed into convex forms, hence the alternative optimization (AO) method can be adopted to decouple the power and bandwidth variables. 
	
	\item Finally, the performance trade-off between S\&C services and that among users are analyzed by numerical results.  
\end{itemize}

The rest of this paper is organized as follows. The mathematical model for the signal transmissions of the proposed ISAC system is established in Section \ref{SystemModel}. In Section \ref{TarDetection}, we formulate the optimization problem of RA for detection service and obtain the associated solutions, after giving the definition of detection QoS. The similar works for localization and tracking services are given in Section \ref{TarLocalization} and Section \ref{TarTracking}, respectively. The simulation results and the trade-off analysis for the S\&C performance are provided in Section \ref{SR}. Finally, we conclude this paper in Section \ref{Conclusion}.      

The notations used in this paper are as follows. Upper-case $\textbf{A}$ (lower-case $\textbf{a}$) bold characters denote matrices (column vectors), and lower case normal letters $a$ are scalars; $(\cdot)^T$, $(\cdot)^*$ and $(\cdot)^H$ represent the transpose, conjugate and complex conjugate transpose operations respectively; $|a|$ stand for the magnitude of a scalar $a$;  $\mathbb{E}\{\cdot \}$ is the statistical expectation; $\textrm{diag}\{\textbf{a}\}$ stands for a diagonal matrix using the elements of $\textbf{a}$ as its diagonal elements; $\textbf{I}_N$ is the $N$-dimensional identity matrix and $\textbf{1}_M$ is the $M \times 1$ vector having all-one entries.

\section{System Model}\label{SystemModel}
\subsection{The General Framework}
As shown in Fig. \ref{system_diagram}, we consider a collocated MIMO BS equipped with $N_t$ transmit and $N_r$ receive antennas in a single-cell setting, providing S\&C services simultaneously in a multi-beam mode. The BS serves $M$ objects in total, which can be classified into the following three types. 
\begin{itemize}
	\item \textit{Sensing targets}: The targets that need to be sensed, e.g., the humans, buildings, monitoring areas, etc, which can be either device-free or device-based. 
	\item \textit{Communication users}: The terminals such as smartphones which require high-quality communication services only.
	\item \textit{ISAC users}: The users that require both S\&C services, i.e., the intersection of sensing targets and communication users. The ISAC users could be people, vehicles, and drones which have communication equipment, namely the device-based targets. 
\end{itemize}

\begin{figure}[!t]
	\centering
	\includegraphics[width=3in]{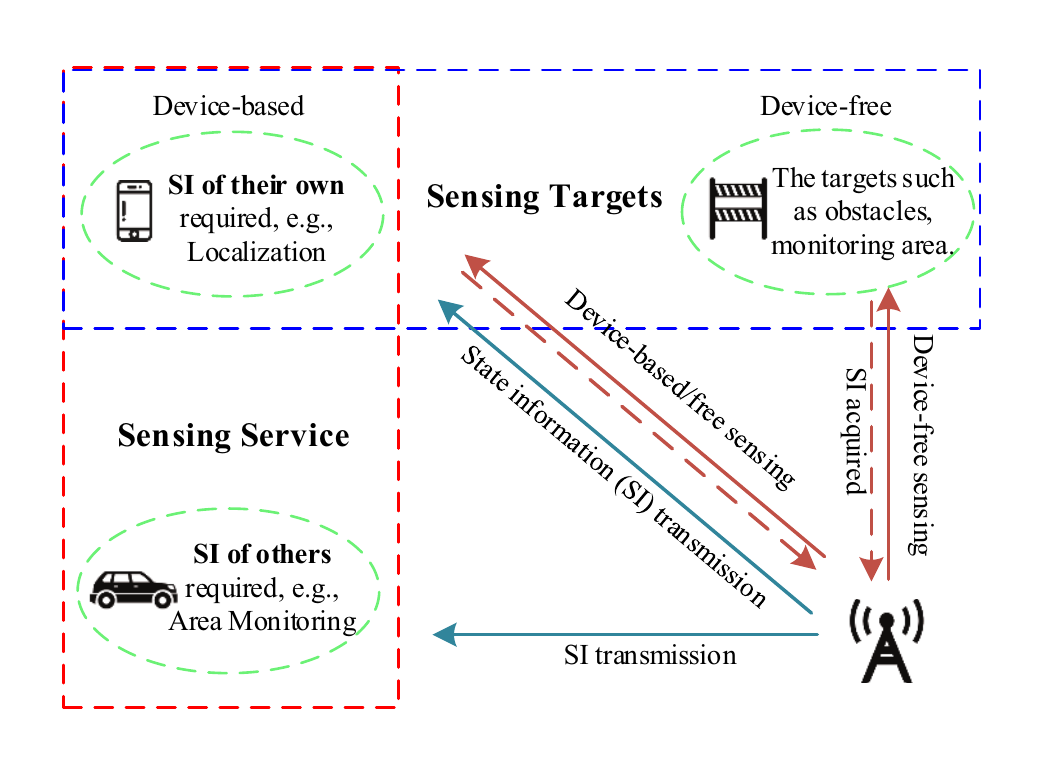}
	\caption{The illustration of sensing service and the targets to be sensed.}
	\label{Sensing_service}
\end{figure}

Intuitively, sensing service can only be provided to device-based users, which can be classified into the following two types as shown in Fig. \ref{Sensing_service}.
	
\begin{itemize}
	\item \textit{The users that require the state information (SI) of their own}: In this case, the targets to be sensed are the users themselves, where a typical application is localization. 
	\item \textit{The users that require the SI of others}: In this case, the BS acquires the SI of other objects via active sensing, and transmits this information through the communication function. A typical use case is vehicle-to-everything (V2X), where the vehicles can sense the obstacles beyond the line of sight through such an ISAC system.
\end{itemize}

\begin{figure}[!t]
	\centering
	\includegraphics[width=3in]{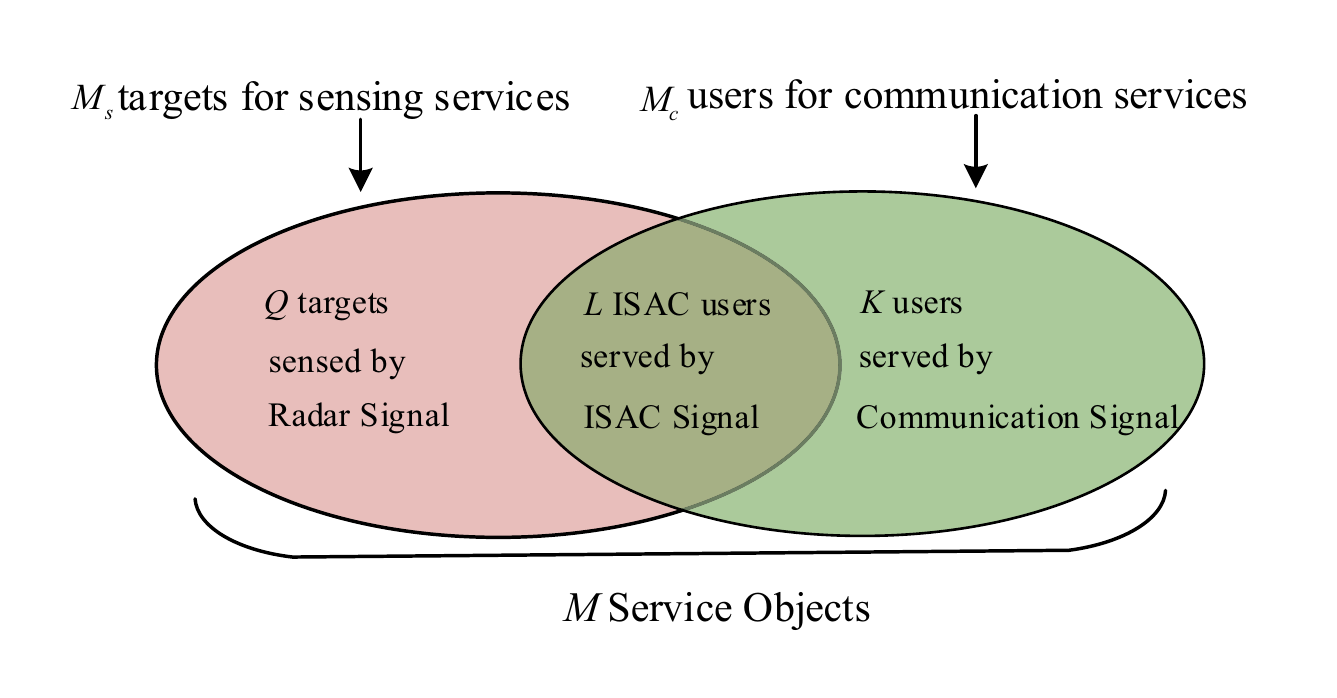}
	\caption{The types of services, users, and transmit signals.}
	\label{RA_Signal_comp}
\end{figure}

Without loss of generality, we assume that the BS transmits radar signals, ISAC signals (also referred to as DFRC signals)  and communication signals correspondingly to $Q$ sensing targets, $L$ ISAC users and $K$ communication users as shown in Fig. \ref{RA_Signal_comp}. For descriptional convenience, we denote the set $\mathcal{I}_s$ with $M_s=Q+L$ elements as the targets to be sensed, and $\mathcal{I}_c$ with $M_c=L+K$ elements as the users which require communication services. Throughout this paper, the system parameters (e.g., transmitted signals) are all constructed in the order of sensing targets, ISAC users, and communication users. Moreover, we also assume that all the users are spatially well-separated so that the main lobes of the beams to the desired directions do not overlap. 

\subsection{Signal Model}
\textit{(1) Transmit Signal Model:} Denote the baseband signals as  $\tilde{\textbf{s}}_n(t)=[\tilde{s}_{1,n}(t),\dots,\tilde{s}_{M,n}(t)]^T \in \mathbb{C}^{M \times 1}$, then the transmitted signal can be expressed as
\begin{equation}
\textbf{s}_n(t)=\textbf{F}_n\tilde{\textbf{s}}_n(t),
\end{equation} 
where $\textbf{F}_n=[\textbf{f}_{1,n},\cdots,\textbf{f}_{M,n}] \in \mathbb{C}^{ N_t \times M}$ is the transmit precoding matrix. The transmitted signals are temporally white wide-sense stationary stochastic processes and statistically independent with zero mean. Accordingly, the transmitted signal covariance is $\textbf{R}_{\textbf{s}_n}=\mathbb{E}\{\textbf{s}_n\textbf{s}^H_n \}=\textbf{F}_n\textbf{F}_n^H$.   

\textit{(2) Target Echo Signal Model:} The baseband representation of the reflected echoes received at the $n$-th epoch can be given in the form
\begin{equation}
\begin{array}{l}\label{ReceiveM}
\textbf{r}_n(t)= \textbf{W}_n^H \sum \limits_{q=1}^{M_s} \kappa \sqrt{p_{q,n}} \alpha_{q,n} e^{\jmath 2 \pi u_{q,n} t} \times \\
\kern 35pt \textbf{a}_r(\theta_{q,n})\textbf{a}^H_t(\theta_{q,n})\textbf{s}_n(t-\tau_{q,n})+\textbf{W}_n^H\textbf{z}_r(t),
\end{array}
\end{equation}
where $\kappa=\sqrt{N_tN_r}$ is the array gain factor. $p_{q,n}$, $\alpha_{q,n}$, $u_{q,n}$, and $\tau_{q,n}$ are the transmit power, the reflection coefficient, the Doppler frequency and the time-delay for the $q$-th target ($q\in\mathcal{I}_s$), respectively. $\textbf{W}_n=[\textbf{w}_{1,n},\cdots,\textbf{w}_{M_s,n}] \in \mathbb{C}^{ N_r \times M_s}$ represents the received beamforming matrix.  $\textbf{z}_r \in \mathbb{C}^{N_r \times 1}$ is the complex additive white Gaussian noise (AWGN) with zero mean and variance of $\sigma_r^2$.

Without loss of generality, the uniform linear array (ULA) with half-wavelength spacing is equipped at the BS, resulting in the following transmit and receive steering vectors 
\begin{equation}
\textbf{a}_t(\theta)=\sqrt{\frac{1}{N_t}}\left[1,e^{\jmath \pi \sin\theta},\cdots,e^{\jmath \pi (N_t-1) \sin \theta}\right]^T,
\end{equation}
\begin{equation}\label{R_vector}
\textbf{a}_r(\theta)=\sqrt{\frac{1}{N_r}}\left[1,e^{\jmath \pi \sin\theta},\cdots,e^{\jmath \pi (N_r-1) \sin \theta}\right]^T.
\end{equation}

\textit{(3) Communication Received Signal Model:} At the $n$-th epoch, the $k$-th user receives the signal from the BS by using a receive beamformer $\tilde{\textbf{w}}_{k,n}$, yielding
\begin{equation}
\begin{aligned}\label{ReceiveCom}
c_{k,n}(t)= \tilde{\kappa}\sqrt{p_{k,n}}\tilde{\alpha}_{k,n}e^{\jmath 2 \pi \varrho_{k,n} t} \times \kern 90pt \\
\tilde{\textbf{w}}_{k,n}^H\textbf{u}(\theta_{k,n})\textbf{a}_t^H(\theta_{k,n})\textbf{f}_{k,n}s_{k,n}(t-\tilde{\tau}_{k,n})+\tilde{\textbf{w}}_{k,n}\textbf{z}_c(t),
\end{aligned}
\end{equation}
where $k \in \mathcal{I}_c$. The steering vector $\textbf{u}(\theta)$ of the user's antenna array has the similar definition as in (\ref{R_vector}) with $\tilde{N}_{r_k}$ antennas. Here, we assume $\tilde{N}_{r_1}=\cdots=\tilde{N}_{r_K}=\tilde{N}_r$ for notational convenience. Again, $\tilde{\kappa}=\sqrt{N_t\tilde{N}_r}$, $\tilde{\alpha}_{k,n}$, $\varrho_{k,n}$ and $\tilde{\tau}_{k,n}$ represent the array gain factor, the communication channel coefficient, the Doppler frequency and the time-delay for the $k$-th user, respectively. To avoid deviation of the core of this paper, we assume that the Doppler frequency and the time-delay can be perfectly compensated through synchronization. Moreover, $\textbf{z}_c(t)$ is the zero-mean AWGN with variance $\sigma_c^2$.   

In this paper, the transmit and receive beamformers can be implemented as
\begin{equation}\label{BF}
\textbf{f}_{i,n}=\textbf{a}_t(\hat{\theta}_{i,n}), \kern 2pt \textbf{w}_{q,n}=\textbf{a}_r(\hat{\theta}_{q,n}), \kern 2pt \tilde{\textbf{w}}_{k,n}=\textbf{u}(\hat{\theta}_{k,n}),
\end{equation}
where $\hat{\theta}_{i,n}, i=1,\cdots,M$ represent the estimate of angle-of-departure (AoD) for the $i$-th user at the $n$-th epoch. Similarly,  $\hat{\theta}_{q,n} (q\in \mathcal{I}_s)$ and $\hat{\theta}_{k,n} (k\in \mathcal{I}_c)$ represent the estimates of AoAs for the sensing targets and communication users, respectively.
 
\subsection{Communication QoS}
In general, communication tasks focus on transmitting information in an efficient and reliable manner. While there are many performance metrics such as spectral efficiency, BER, etc., we choose the commonly employed achievable rate to measure the communication QoS, since this metric directly depends on the cost of wireless resources, e.g., power and bandwidth. As we allocate orthogonal frequency bandwidth to different users, the inter-user interference can be omitted at the communication receiver. Thus, the received SNR for the $k$-th user at the $n$-th epoch can be obtained by
\begin{equation}
\begin{aligned}
\text{SNR}_{k,n}
&=\frac{p_{k,n}|\tilde{\kappa}\tilde{\alpha}_{k,n}(\tilde{\textbf{w}}_{k,n})^H\textbf{u}(\theta_{k,n})\textbf{a}_t^H(\theta_{k,n})\textbf{f}_{k,n}|^2}{{b_{k,n}}\sigma_c^2}\\
&=p_{k,n}\varsigma_{k,n}^c/b_{k,n}, \\
\end{aligned}
\end{equation}
where 
\begin{equation}
\varsigma_{k,n}^c=|\tilde{\kappa}\tilde{\alpha}_{k,n}(\textbf{w}_{k,n}^c)^H\textbf{u}(\theta_{k,n})\textbf{a}_t^H(\theta_{k,n})\textbf{f}_{k,n}|^2/\sigma_c^2,
\end{equation}
is defined as the communication channel gain normalized by the noise power. The achievable sum-rate of all $M_c$ communication users can be expressed by
\begin{equation}\label{Sumrate}
R_n(\textbf{p}_{\mathcal{I}_c},\textbf{b}_{\mathcal{I}_c})=\sum \limits_{k=1}^{M_c} b_{k,n} \log_2\left( 1+\frac{p_{k,n}\varsigma_{k,n}^c}{b_{k,n}}\right) , 
\end{equation}
where $b_{k,n}$ is the bandwidth allocated to the $k$-th user. $\textbf{p}_{\mathcal{I}_c}$ and $\textbf{b}_{\mathcal{I}_c}$ are the power and bandwidth allocation vectors for all $M_c$ users.  

\subsection{A Unified Resource Allocation Framework}
In the proposed ISAC RA framework, the total system power $P_\text{total}$ and frequency bandwidth $B_\text{total}$ need to be properly allocated among the $M$ objects to meet different QoS demands. The BS transmitter exploits a portion of the total system resources to sense targets, whereas the other portion is employed for information transmission. Our aim is optimizing the sensing QoS for various tasks under the constraints of communication QoS requirement and the resource limitation, to reveal the performance trade-off between S$\&$C services associated with the power and bandwidth resources. Therefore, the optimization problem for a unified RA framework can be constructed as following  
\begin{equation}\label{unified}
\begin{aligned}
	\mathop { \text{maximize} }\limits_{\textbf{p},\textbf{b}} \kern 10pt & \text{Sensing QoS (detection, localization, tracking)}   \\
	\text{subject to}  \kern 10pt & R(\textbf{p},\textbf{b})\ge \Gamma_c, \kern 2pt \textbf{1}_{M}^T\textbf{p}=P_\text{total},\kern 2pt \textbf{1}_{M}^T\textbf{b}=B_\text{total}.
\end{aligned}
\end{equation}
In the next three sections, we first define the sensing QoS for detection, localization and tracking, respectively. Then, the corresponding RA algorithms are proposed to obtain the optimal/sub-optimal solutions for problem (\ref{unified}).

\section{Target Detection}\label{TarDetection}
The detection task is to determine the existence of a target in the area of interest, of which applications include unlicensed UAV monitoring, traffic accident area monitoring, etc. High detection QoS is required for sensing scenarios where missing alarms may cause severe consequences. In this section, we give the definition of the detection QoS and formulate the RA problem based on different service requirements. We tentatively drop off the epoch index $n$ since the target detection should be implemented at each epoch individually.   
\subsection{The Detection QoS}\label{TDQos}
In general, radar systems transmit omni-directional waveforms for the initial detection. In this stage, the system resources are uniformly distributed since the resource allocation does not make sense without any prior information. However, the omni-directional detection scheme is not suitable for the proposed urban perceptive network due to the following reasons: 1) The complex surroundings lead to a large amount of clutter; 2) There may be no detection service requirements in the majority of directions. Therefore, the detection task with a specific purpose is considered in this section.

Specifically, let us consider the following two application scenarios: 1) We have the target information at the $(n-1)$-th epoch and need to detect its existence at the $n$-th epoch; 2) We do not have any information but wish to detect a specific area with the given $\hat{\theta}_q$ and $\hat{\sigma}_{\alpha_q}$. Based on the above discussion, we assume that the prior information on angle $\hat{\theta}_q$ and the standard deviation of reflection coefficient $\hat{\sigma}_{\alpha_q}$ for the $q$-th target are known to the BS. Here, $\hat{\sigma}_{\alpha_q}$ can be determined by the desired distance and radar cross section (RCS) of the target based on radar equation \cite{RadarSP}. 

The BS can implement beamformer (\ref{BF}) by exploiting the prior angle information, where the beamforming gain factor can be calculated by
\begin{equation}\label{beamerror}
\epsilon_q=|\textbf{a}_t^H(\theta_q)\textbf{a}_t(\hat{\theta}_q)|, \kern 4pt \tilde{\epsilon}_q=|\textbf{a}_r^H(\theta_q)\textbf{a}_r(\hat{\theta}_q)|,
\end{equation}    
where we have $\epsilon_q,\tilde{\epsilon}_q \in [0,1]$. Then, the spatial-temporal matched filter output of received signal (\ref{ReceiveM}) in a given delay-Doppler $(\tau_q,u_q)$ bin can be recast to    
\begin{equation}\label{GRD}
x = \sqrt{p_q} \alpha_q \kappa^2\epsilon_q \tilde{\epsilon}_q +z_q,
\end{equation}  
where $x=\int r_{q,n}(t)s_q^*(t-\tau_q) e^{-\jmath 2 \pi u_q t} dt$, $r_{q,n}(t)$ is the $q$-th element of $\textbf{r}_n(t)$. $z_q=\int \textbf{w}_q^H\textbf{z}_r(t)s_q^*(t-\tau_q)e^{-\jmath 2 \pi u_q t}dt$ is the match filtering noise, which follows $z_q \sim \mathcal{CN}(0,N_r\sigma_r^2)$. The term $\kappa^2$ denotes the product of array gain and beamforming gain. Here, we assume that the targets do not share the same range-Doppler bin if they locate in the same beam. This assumption implies that different targets can always be distinguished by either beamformer or matched filter. 

The target detection problem can be formulated as the following composite binary hypothesis test \cite{RadarSP}   
\begin{equation}
\begin{array}{l}
\mathcal{H}_0 : \text{No target in the delay-Doppler bin} \kern 2pt (\tau,u)\\
\mathcal{H}_1 : \text{Target exists in the delay-Doppler bin} \kern 2pt (\tau,u).
\end{array}
\end{equation}
 Accordingly, the detection model can be described by the following hypothesis testing problem \cite{FE2006}
\begin{equation}\label{Decision}
x_q=\left\{
\begin{array}{l}
\mathcal{H}_0 : z_q\\
\mathcal{H}_1 : \sqrt{p_q}\alpha_q\kappa^2\epsilon_q \tilde{\epsilon}_q+z_q,
\end{array}\right.
\end{equation}
where the channel reflection coefficient is a complex Gaussian random variable with $\alpha_q \sim \mathcal{CN}(0,\hat{\sigma}_{\alpha_q}^2)$ for the general case \cite{FE2006}. The optimal detector is given by \cite{BookDetectionTheory}       
\begin{equation}\label{eee}
E=|x|^2\mathop \lessgtr \limits^{\mathcal{H}_0}_{\mathcal{H}_1} \delta,
\end{equation}
where the decision threshold $\delta$ is set to meet the desired false alarm rate. It is easy to see the test statistic $E$ subjects to the following distribution \cite{FE2006}
\begin{equation}
E \sim \left\{
\begin{aligned}
&\frac{N_r\sigma_r^2}{2}\chi_2^2, &\mathcal{H}_0 \\
&\left( \frac{N_r\sigma_r^2}{2}+\frac{p_q\hat{\sigma}_{\alpha_q}^2\kappa^4\epsilon_q^2\tilde{\epsilon}_q^2}{2}\right)\chi_2^2,\kern 5pt  &\mathcal{H}_1  
\end{aligned} \right.,
\end{equation}
where $\chi_2^2$ is the central chi-squared distribution with two degrees of freedom. For a given constant $P_\text{FA}$, i.e., $P_{\text{FA}}=Pr(\chi_2^2>2\delta/N_r\sigma_r^2)$, the decision threshold $\delta$ in (\ref{eee}) can be determined by
\begin{equation}
\delta = \frac{N_r\sigma_r^2}{2}F_{\chi_2^2}^{-1}(1-P_{\text{FA}}).
\end{equation} 
Then, the probability of detection can be determined by
\begin{equation}\label{PD}
P_\text{D}=Pr(E>\delta|\mathcal{H}_1)=1-F_{\chi_2^2}\left( \frac{2\delta/(N_r\sigma_r^2)}{1+p_q\varsigma_q}\right),
\end{equation}  
where $F_{\chi_2^2}(\cdot)$ and $F_{\chi_2^2}^{-1}(\cdot)$ are the cumulative distribution function (CDF) of the chi-square random variable and its inverse operation, respectively. Here, the normalized sensing channel gain is defined as
\begin{equation}\label{CGN}
\varsigma_q = \frac{\hat{\sigma}_{\alpha_q}^2\kappa^4\epsilon_q^2\tilde{\epsilon}_q^2}{N_r\sigma_r^2}.
\end{equation} 
We note that there are various detection models that differ by the assumed signal model, and the unknown parameters, etc. We refer the readers to \cite{FE2006} and \cite{IB2006} for details.

\subsection{Problem Formulation}
Our goal is to maximize the detection QoS under the constraint of communication QoS by a sophisticatedly tailored RA scheme. It can be observed in (\ref{PD}) that $P_\text{D}$ is a monotonically increasing function with respect to (w.r.t.) the parameter $\rho_q=p_q\varsigma_q$ according to the property of CDF function. Thus, we relax the problem of maximizing $P_\text{D}$ into a more tractable problem of maximizing $\rho_q$. Typically, problem (\ref{unified}) can be transformed into the following two forms based on different criteria.

(1) \textit{Fairness}: One can maximize the minimum $P_\text{D}$ of each sensing target to guarantee the fairness, then the power allocation problem under this criterion can be formulated as
\begin{equation}\label{ProDetectFair}
\begin{aligned}
\mathop { \text{maximize} }\limits_{\textbf{p}} \kern 10pt &F(\textbf{p})= \text{min}\{\rho_1,\cdots,\rho_{M_c}\}   \\
\text{subject to}  \kern 10pt &\textbf{1}_{M}^T\textbf{p}=P_\text{total},\kern 2pt R(\textbf{p})\ge \Gamma_c, \\
&P_\text{min} \le p_i \le P_\text{max}, i=1,\cdots,M, \\ 
\end{aligned}
\end{equation}  
where $\textbf{p}=[p_1,\cdots,p_M]$ is the power vector to be allocated for the transmit signals. The first two constraints represent the total power budget and the communication QoS requirement which ensures that the system sum-rate is larger than a certain threshold $\Gamma_c$. The last constraint imposed is to avoid the extremely large/small power allocation, guaranteeing the basic system functions. Problem (\ref{ProDetectFair}) is convex w.r.t. variable $\textbf{p}$, hence the off-the-shelf CVX toolbox can be employed to obtain the optimal solution efficiently \cite{CVX}. Furthermore, it should be noted that (\ref{ProDetectFair}) is equivalent to the original problem of maximizing $F(P_\text{D})$ in this case.

(2) \textit{Comprehensiveness}: To achieve the optimal system's sensing performance, the objective can be formulated as the sum $P_\text{D}$ of all targets. Similarly, the power allocation problem is expressed as
\begin{equation}\label{ProDetect}
\begin{aligned}
\mathop { \text{maximize} }\limits_{\textbf{p}} \kern 10pt &F(\textbf{p})=\sum \limits_{q=1}^{M_s} \rho_q \\
\text{subject to}  \kern 10pt &\textbf{1}_M^T\textbf{p}=P_\text{total}, \kern 2pt R(\textbf{p})\ge \Gamma_c, \\
&P_\text{min} \le p_i \le P_\text{max}, i=1,\cdots,M, \\ 
&\rho_1:\cdots:\rho_{M_s}=\gamma_1:\cdots:\gamma_{M_s}. 
\end{aligned}
\end{equation} 
Inspired by \cite{Propor2005}, the proportional rate constraints imposed are to ensure the proportional fairness among sensing targets that have different importance levels. These constraints can be omitted if there is no specific importance requirement. Additionally, it is reported in \cite{Propor2005} this problem is equivalent to the \textit{max-min} problem when all $\{\gamma_i\}_{i \in \mathcal{I}_s}$ are equal. Interestingly, this equivalence between problem (\ref{ProDetectFair}) and (\ref{ProDetect}) does not always hold due to the existence of ISAC users, which depends on the value of $\Gamma_c$. When the constraints of proportional fairness and the achievable sum-rate are contradictory to each other for large $\Gamma_c$, the sensing performance among the users achieved by solving (\ref{ProDetectFair}) is no longer the same whereas the problem (\ref{ProDetect}) becomes infeasible. We will also show this phenomenon in Section \ref{SR} by numerical simulation. Again, problem (\ref{ProDetect}) is still convex and can be readily solved.    

\section{Target Localization}\label{TarLocalization}

The localization task is to estimate the range and azimuth angle of the target, thereby determining the location in space. Usually, nearby targets with large sizes require relatively low localization accuracy, or equivalently, lower localization QoS than the far and small ones. Moreover, we usually prefer to allocate more resources to the important target than the others. Motivated by this, we define the localization QoS and formulate the joint power and bandwidth allocation problem to ensure both performance of S\&C functionalities in this section.

\subsection{The Localization QoS}
The radar estimation parameters include range $d$ (time delay), velocity $v$ (Doppler frequency), and azimuth angle $\theta$ (AoA/AoD). To assess the performance of an estimator, the MSE is a straightforward metric, i.e., $\mathbb{E}\{| \bm{\phi}-\hat{\bm{\phi}}|^2 \}$, where $\bm{\phi}=[d,v,\theta]$ represents the parameter vector. However, a closed-form MSE is difficult to characterize for system design. Alternatively, CRB is adopted in many applications as a theoretical limit for the MSE. The exact expressions of CRB relies on the system configuration and signal structures, which have various forms \cite{liu2021survey}. To simplify the expression, we adopt the generalized form to characterize the estimation CRBs for the $q$-th sensing target as \cite{ZHW2020,Yan2015,Yan2017cooperative}
\begin{equation}\label{CRB}
	\left\{
	\begin{array}{l}
		\text{CRB}(d_q) \propto (p_{q}|\varsigma_{q}|^2B_\text{rms,q}^2)^{-1}   \\
		\text{CRB}(v_q) \propto (p_{q}|\varsigma_{q}|^2T_\text{e,q}^2)^{-1}   \\
		\text{CRB}(\theta_q) \propto (p_{q}|\varsigma_{q}|^2/\text{W}_\text{NN})^{-1}  
	\end{array}\right. ,
\end{equation}
where $\text{W}_\text{NN}$ is null-to-null beam width of the receive antenna, $T_\text{e}$ represents the effective time duration. $B_\text{rms}$ is the effective bandwidth with the expression of 
\begin{equation}\label{BRMS}
	B_\text{rms}^2=\frac{\int_{B} f^2 \left| {S(f)} \right|^2 df}{\int_{B} \left| {S(f)} \right|^2 df},
\end{equation}
where $S(f)$ represents the Fourier transform of the time-domain waveform $\tilde{s}(t)$.

It is shown in (\ref{BRMS}) the effective bandwidth $B_\text{rms}$ relates to the sensing performance through both the normal bandwidth $B$ and frequency-domain waveform $S(f)$, whereas the communication performance depends only on $B$. To unify the unit of bandwidth resource, we adopt a general signal spectrum with the form $S(f)=\sin \pi f T_p/\pi f$, of which bandwidth is limited to a finite value $B$. This is equivalent to passing a perfectly rectangular pulse through a filter of width $B$. By substituting $S(f)$ into (\ref{BRMS}), a reasonably good approximation for almost any value of $B_\text{rms}^2$ can be obtained as \cite{MS1960}
\begin{equation}\label{Bappro}
	B_\text{rms}^2 \approx B/(2\pi^2T_\text{pulse}),
\end{equation}    
where $T_\text{pulse}$ represents the pulse width. Note that the general relationship between the effective and the normal bandwidth can be obtained by the similar process, once the time-domain waveform is given. As an example, for the linear frequency modulation wave, we have $B_\text{rms}^2 \approx B^2/6$. The readers are referred to \cite{MS1960} for more details.

\subsection{Motivation on Bandwidth Allocation} 

The motivation of power allocation is straightforward since the power resource has similar impacts on S\&C performance. By contrast, the impacts of bandwidth resource imposed on radar and communication systems are quite different. As shown in Fig. \ref{BAD}, let us consider the following two types of interference.

\textit{(1) Interference between S\&C services in a single beam.} In this case, orthogonal bandwidth should be allocated among the communication users to avoid inter-user interference since the communication users are able to operate on heterogeneous frequency bands. By contrast, sensing targets will reflect the signals at all frequency bands. Hence, all the sensing targets in a single beam have to share the same bandwidth. Nevertheless, non-overlapped bandwidth should be allocated between S\&C services to circumvent the interference of radar signals to communication users.
  
\textit{(2) Inter-beam interference among multiple beams.} In different beams, the interference for S\&C both stems from the power leakage from one beam to another, i.e., the overlapped main-lobes or the side-lobes. Obviously, the inter-beam interference can be alleviated by employing the matched filtering method with orthogonal bandwidth. However, it should be highlighted that the beam formulated will be sufficiently narrow by scaling up the number of antennas, such that the inter-beam interference can be omitted \cite{ngo2015massive}. In such a scheme, allocating the maximum available bandwidth to each target and user is more reasonable for improving the system performance rather than the orthogonal bandwidth allocation, which is however at the price of enlarged antenna array.   

 \begin{figure}[!t]
	\centering
	\includegraphics[width=3.5in]{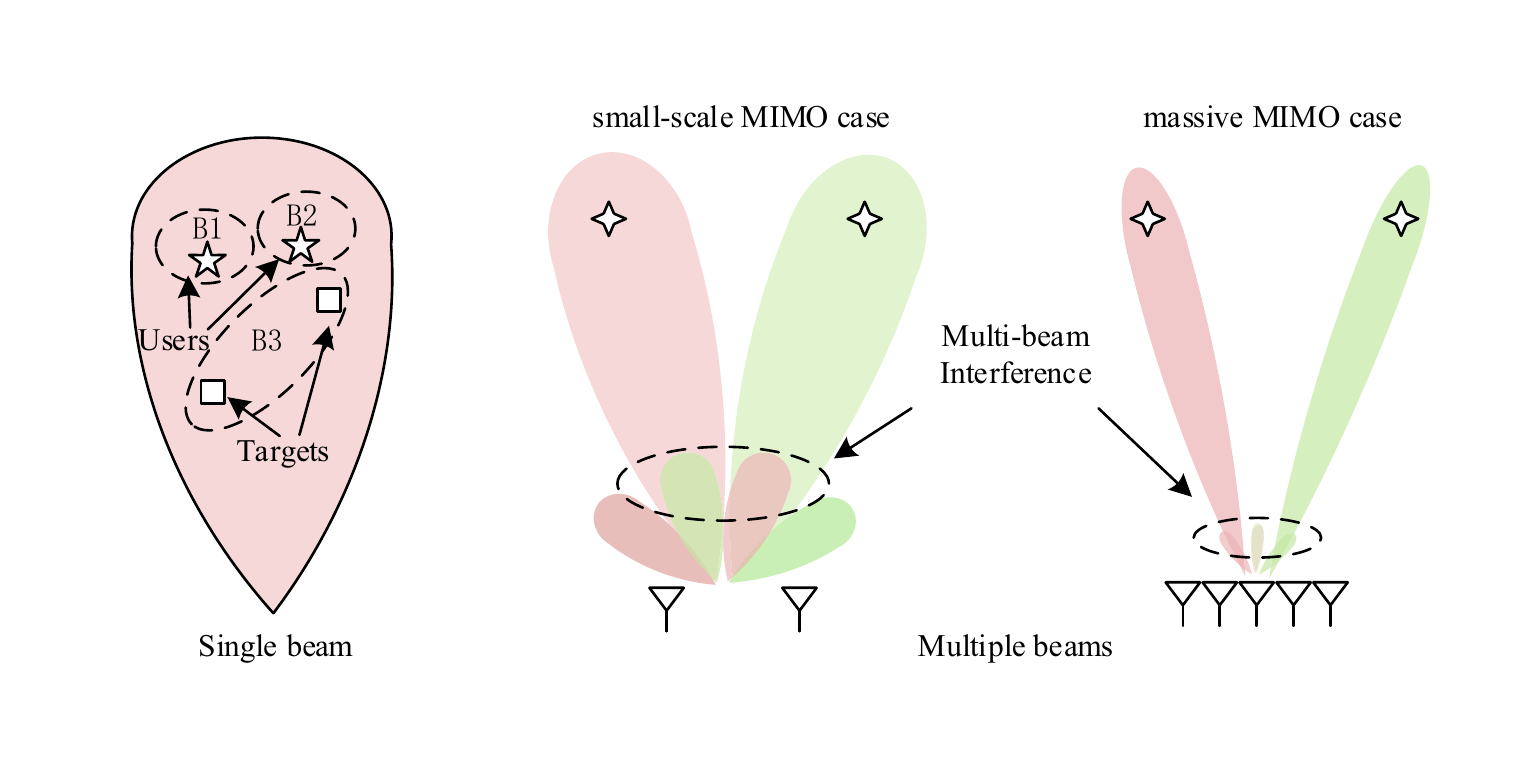}
	\caption{Bandwidth allocation in different scenarios.}
	\label{BAD}
\end{figure}

By substituting (\ref{Bappro}) into (\ref{CRB}), it is shown that the CRB of range estimation is inversely proportional to the product of the received SINR and bandwidth, which is referred to as the SBP. Similarly, a large SBP results in a high achievable sum-rate for communication systems. As per the above analysis, the bandwidth and the interference determining the SINR are coupled in the SBP. Namely, allocating the total bandwidth resource for all targets will introduce multi-beam interference, resulting in a degraded SINR level and an indeterminate SBP. In what follows, we will evaluate the trade-off between the bandwidth and the spatial/hardware resources by comparing the SBP, to provide insights into the choice of the optimal bandwidth allocation strategy with a fixed number of antennas.

The received signal for the $q$-th target/user with perfect beamforming can be expressed as
\begin{equation}\label{rq}
\begin{aligned}
r_q(t)=\underbrace {\kappa^2\textbf{a}_r^H(\theta_q)\textbf{a}_r(\theta_q)\textbf{a}_t^H(\theta_q)\textbf{a}_t(\theta_q)s_q(t)}_{\text{Desired signal}}
+\kern 40pt\\
\underbrace {\kappa^2\textbf{a}_r^H(\theta_q)\sum\limits_{i = 1}^M\sum\limits_{j=1,j\ne q}^M\textbf{a}_r(\theta_i)\textbf{a}_t^H(\theta_i)\textbf{a}_t(\theta_j)s_j(t)+z(t)}_{\text{Interference plus noise signal}}.
\end{aligned}
\end{equation} 
By recalling (\ref{beamerror}), we replace the estimation angle $\hat{\theta}_q$ by the other user's AoA $\theta_i$, then $\epsilon_{q,i}$ and $\tilde{\epsilon}_{q,i}$ represent the inter-beam interference factors as well. For convenience, we only keep the first-order terms w.r.t. $\epsilon_{q,i}$ and $\tilde{\epsilon}_{q,i}$, namely the terms with $i=q$ and $i=j \ne q$. 

For the scheme using maximum bandwidth without bandwidth allocation, the SBP can be expressed as 
\begin{equation}\label{SBP1}
\text{SBP}_\text{total}=\frac{\kappa^2}{\kappa^2\sum\limits_{i = 1}^M(\epsilon_{q,i}+\tilde{\epsilon}_{q,i})+\sigma_\text{eff}^2}B_\text{total}.
\end{equation}  
where $\sigma_\text{eff}^2$ is the equivalent noise power by taking the thermal noise, environmental clutter, and other relevant interference into account. By contrast, when allocating orthogonal bandwidth to different users, the first-order interference can be eliminated by matched filtering. In this scheme, the SBP can be given by
\begin{equation}\label{SBP2}
\text{SBP}_\text{orth}=\frac{\kappa^2}{\sigma_\text{eff}^2}b_q.
\end{equation}    
where $b_q$ is the bandwidth allocated to the $q$-th user. Then, the SBP ratio of the above two schemes can be obtained by
\begin{equation}\label{SBPratio}
\frac{\text{SBP}_\text{total}}{\text{SBP}_\text{orth}}= \frac{\sigma_\text{eff}^2}{\kappa^2\sum\limits_{i = 1}^M(\epsilon_{q,i}+\tilde{\epsilon}_{q,i})+\sigma_\text{eff}^2} \frac{B_\text{total}}{b_q}.
\end{equation} 
On the right side of this equation, $B_\text{total}/b_q$ is always larger than 1 but the front fraction is less than 1. It implies that there will be an equilibrium point between these two schemes (i.e., the ratio is equal to 1). As shown in Fig. \ref{SBP}, the SBP ratio increases upon the increasing number of antennas which formulates narrower beam. Furthermore, we have $\epsilon_{q,i}, \tilde{\epsilon}_{q,i} \to 0$, when $N_t,N_r \to \infty$ \cite{ngo2015massive}. Therefore, transmitting maximum bandwidth is superior to the orthogonal bandwidth allocation scheme in the massive MIMO scheme. However, orthogonal spectrum allocation is necessary for small-scale MIMO cases, especially in the high SNR regime. Overall, we allocate the non-overlapped bandwidth to all S\&C users to avoid interference in this paper. 

 \begin{figure}[!t]
	\centering
	\includegraphics[width=3in]{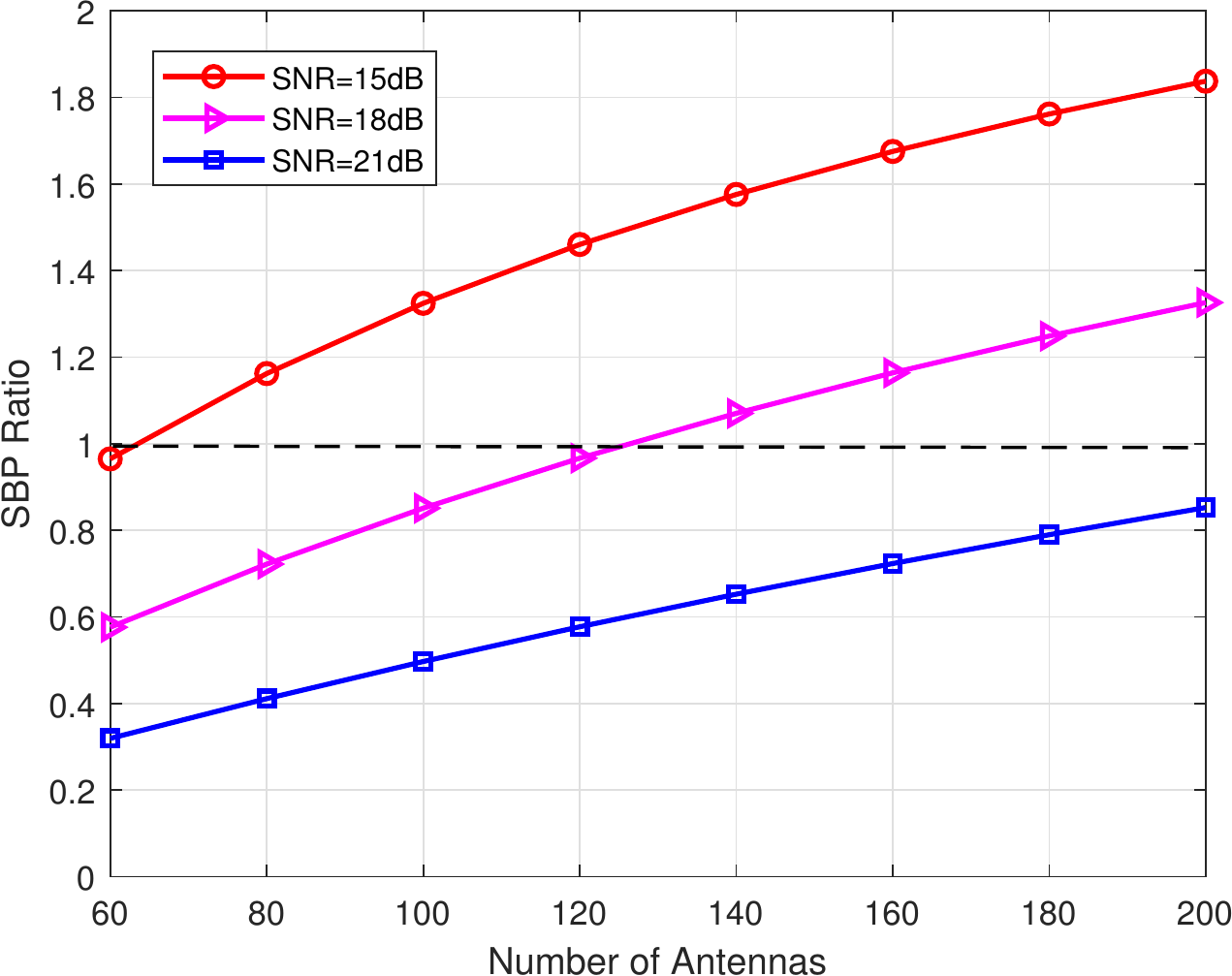}
	\caption{The illustration of the trade-off between the bandwidth and spatial resources. The AoAs of three user are $[-30^\circ, 0^\circ, 30^\circ]$, where the reference user locates at $\theta=0^\circ$. SNR is defined as $10\log(\kappa^2/\sigma_\text{eff}^2)$, and $B_q=1/3B_\text{total}$.  }
	\label{SBP}
	\setcounter{figure}{4}
\end{figure}
\textbf{Remark:} Note that there is no longer interference between sensing target and communication user by transmitting ISAC signals. To be specific, the communication interference in the echo signal can be removed since the BS has prior knowledge on its transmitted symbols. This implies that the spectrum can be overlapped between sensing target and communication user if the BS transmits ISAC signals only. While the use of ISAC signals may increase the interference-free bandwidth, dedicated radar/communication signals are expected to achieve better performance than ISAC signal, as they are specifically tailored for individual functionalities rather than for simultaneous S\&C transmission. Accordingly, there is another performance trade-off between performance gain and spectral resource consumption. We will leave this as an open problem for future research.
        
\subsection{Problem Formulation}\label{RAL} 
Localization focuses on the distance and AoA estimation for a static target, where the time index $n$ can also be omitted. Our goal is to minimize the localization error, i.e., the CRBs of distance and AoA, while guaranteeing the communication QoS. For convenience, we assume that      
\begin{equation}\label{crb}
	\text{CRB}(d_{q})= \frac{\beta_1}{p_{q}|\varsigma_{q}|^2b_{q}},
	\text{CRB}(\theta_{q}) = \frac{\beta_2}{p_{q}|\varsigma_{q}|^2},
\end{equation} 
where the scale factors $\beta_i, i=1,2$ introduced are constants related to the system configuration, signal designs, beamforming gain, matched filtering gain as well as the specific signal processing algorithms\cite{liu2020radar}. We equivalently transform the minimization problem of CRB into the maximization problem of its reciprocal, where the objective parameter for the $q$-th sensing target can be expressed as 
\begin{equation}
\rho_{q}=  \frac{\omega_{\tau}}{\text{CRB}(d_{q})} + \frac{\omega_{\theta}}{\text{CRB}(\theta_{q})} ,
\end{equation}
where $\omega_{\tau}$ and $\omega_{\theta}$ are the normalized factors to unify the units of distance and AoA. Consequently, the optimization problem based on \textit{comprehensiveness} criterion can be formulated as    
\begin{subequations}\label{ProLocal}
\begin{align}
\mathop { \text{maximize} }\limits_{\textbf{p},\kern 2pt \textbf{b}} \kern 10pt & F(\textbf{p},\textbf{b}) =\sum \limits_{q=1}^{M_s} \rho_q  \\
\text{subject to} \kern 10pt & R(\textbf{p},\textbf{b}) \ge \Gamma_c, \label{PLCC} \\
&\textbf{1}_{M}^T\textbf{p}=P_\text{total}, \kern 2pt \textbf{1}_{M}^T\textbf{b}=B_\text{total}, \\
&P_\text{min} \le p_{i} \le P_\text{max}, i=1,\cdots,M, \label{PLC1} \\
&B_\text{min} \le b_{j} \le B_\text{max}, j=1,\cdots,M, \label{PLCB} \\
&\rho_{1}:\cdots:\rho_{M_s}=\gamma_1:\cdots:\gamma_{M_s}. \label{PLC2}
\end{align}
\end{subequations}
where $\textbf{p}=[p_{1},\cdots,p_{M}]$ and $\textbf{b}=[b_{1},\cdots,b_{M}]$ are the power and bandwidth allocation vectors. The formulation base on the \textit{fairness} criterion is omitted to avoid repetition. The corresponding \textit{max-min} problem can be readily obtained as similar to (\ref{ProDetectFair}).

Different from the RA problem for detection, the S\&C QoS is related to both power and bandwidth resources for localization. Accordingly, the variables $\textbf{p}$ and $\textbf{b}$ are coupled in both objective function and the constraints (\ref{PLCC}) and (\ref{PLC2}), leading to the non-convexity of problem (\ref{ProLocal}). To decouple the variables, we adopt the alternative optimization (AO) method to obtain the sub-optimal solution. Specifically, for a given initial bandwidth allocation $B$, the sub-problem of (\ref{ProLocal}) w.r.t. variable $\textbf{p}$ is obvious convex, where the solution can be obtained by CVX effectively. Subsequently, for the fixed power allocation $\textbf{p}$, the corresponding sub-problem of (\ref{ProLocal}) w.r.t. variable $\textbf{b}$ is again convex. It is because that the achievable sum-rate (\ref{Sumrate}) is a concave function w.r.t. $\textbf{b}$ such that the constraint (\ref{PLCC}) is convex (the proof can be seen in \cite{Propor2005}, Appendix I). Moreover, the other constraints and the objective are all linear. Thus the variable $\textbf{b}$ can be updated by solving this sub-problem. Finally, the optimal solution can be obtained by optimizing $\textbf{p}$ and $\textbf{b}$ alternatively until convergence.     

\textit{Initial Bandwidth}: Note that the performance of AO method depends on the choice of the initial value. Inappropriate initial bandwidth (e.g., uniformly distributed) will lead to an infeasible problem at the first iteration with large $\Gamma_c$. To this end, we propose an initial bandwidth selection method, which is summarized in \textbf{Algorithm} \ref{alg1} together with the proposed joint power and bandwidth allocation algorithm. The main idea is to calculate the maximum achievable sum-rate with the parameter settings and to extract the corresponding bandwidth allocation vector. 

\begin{algorithm}[!t]
	\caption{Joint Power and Bandwidth Allocation}   
	\label{alg1}
	\begin{algorithmic}		
		\STATE \textbf{Input:}$P_\text{total}$, $B_\text{total}$, $\Gamma_c$, $\epsilon$, $\varsigma_{k}^c$, $k\in \mathcal{I}_c$, $\varsigma_q$, $\gamma_q$, $q\in \mathcal{I}_s$ 
		\STATE \textbf{Output:} The power and bandwidth allocation $\textbf{p}$ and $\textbf{b}$ 	
		\STATE {1.} Allocate bandwidth as $b_i=B_\text{total}/M$ uniformly \\
		\STATE {2.} Compute the power allocation as following
		\begin{equation*} \textbf{p}^{(0)}=\left\{
		\begin{aligned}
		\mathop { \text{maximize} }\limits_{\textbf{p}} \kern 10pt & R(\textbf{p},\textbf{b}) \\
		\text{subject to}\kern 10pt & \textbf{1}_{M}^T\textbf{p}=P_\text{total}, \kern 2pt \text{(\ref{PLC1})}, \kern 2pt \text{(\ref{PLC2})} \\
		\end{aligned}\right.
		\end{equation*}
		\STATE {3.} Compute the initial bandwidth allocation
		\begin{equation*} \textbf{b}^{(0)}=\left\{
		\begin{aligned}
		\mathop { \text{maximize} }\limits_{\textbf{b}} \kern 10pt & R(\textbf{p}^{(0)},\textbf{b}) \\
		\text{subject to}\kern 10pt & \textbf{1}_{M}^T\textbf{b}=B_\text{total}, \kern 2pt \text{(\ref{PLCB})}, \kern 2pt \text{(\ref{PLC2})} \\
		\end{aligned}\right.
		\end{equation*}
		\STATE \textbf{while} $|F(\textbf{p}^{(k+1)},\textbf{b}^{(k+1)})-F(\textbf{p}^{(k)},\textbf{b}^{(k)})|>\epsilon$ \textbf{do} \\
		
		\STATE {4.} Compute $\textbf{p}^{(k+1)}$ by solving (\ref{ProLocal}) with fixed $\textbf{b}^{(k)}$	
		\STATE {5.} Compute $\textbf{b}^{(k+1)}$ by solving (\ref{ProLocal}) with fixed $\textbf{p}^{(k+1)}$\\
		\textbf{end while}			
	\end{algorithmic}
\end{algorithm}

\section{Target Tracking}\label{TarTracking}
The tracking task is the extended application of localization. Besides the estimation of the target's state parameters, the state prediction should be considered to improve the tracking performance. In this section, we establish a closed loop tracking system based on the Kalman filter, where the power and bandwidth resources are jointly allocated to guarantee the satisfactory S\&C QoS in each epoch. We consider the constant-velocity targets in state models for simplicity. The proposed approach can readily be extended to other state evolution models by replacing the motion transition matrix.   
  
\subsection{The tracking QoS}
For a moving target, the motion state vector of the $q$-th target at $n$-th epoch as $\bm{\xi}_{q,n}=[x_{q,n},y_{q,n},\dot{x}_{q,n}, \dot{y}_{q,n}]^T$, where $(x_{q,n},y_{q,n})$ and $(\dot{x}_{q,n}, \dot{y}_{q,n})$ denote the position and velocity component in the Cartesian coordinate, respectively. Target tracking is to determine the motion state by the estimated parameters (measurement) and the state prediction.   

\textit{State Evolution Model:} Let us consider the constant-velocity model as following \cite{Yan2015}

\begin{equation}
\bm{\xi}_{q,n} = \textbf{F}_\xi \bm{\xi}_{q,n-1}+\textbf{w}_{q,n-1},
\end{equation} 
where $\textbf{w}_{q,n-1}$ represents the process noise, which is assumed to be zero-mean Gaussian distributed with a known covariance $\bm{\Phi}_q$ as
\begin{equation}
\bm{\Phi}_q = \left[ {\begin{array}{*{20}{c}}
	{\frac{1}{3}T_s^3}&{\frac{1}{2}T_s^2}\\
	{\frac{1}{2}T_s^2}&{{T_s}}
	\end{array}} \right] \otimes {\tilde{\sigma}_q}{\textbf{I}_2},
\end{equation} 
where $\tilde{\sigma}_q$ is the process noise intensity and $T_s$ is the sample interval. $\textbf{F}_\xi$ denotes the transition matrix as
\begin{equation}
\textbf{F}_\xi = \left[ {\begin{array}{*{20}{c}}
	1&T_s\\
	0&1
	\end{array}} \right] \otimes{\textbf{I}_2}.
\end{equation} 

\textit{Measurement Model:} The nonlinear measurement model of the $q$-th target can be described as \cite{ZHW2020}
\begin{equation}
\textbf{y}_{q,n} = h(\bm{\xi}_{q,n})+\tilde{\textbf{w}}_{q,n},
\end{equation} 
where $\textbf{y}_{q,n}=[d_{q,n},v_{q,n},\theta_{q,n}]^T$ represents the vector of radar estimation parameters. Thus, the nonlinear operator $h(\cdot)$ is indeed the coordinate transformation from the Cartesian system to polar system, which is in the form of
\begin{equation}
h(\bm{\xi}_{q,n})=
\left\{
\begin{aligned}
&d_{q,n}=\sqrt{x_{q,n}^2+y_{q,n}^2},   \\
&v_{q,n}=(\dot{x}_{q,n}x_{q,n}+\dot{y}_{q,n}y_{q,n})/d_{q,n}, \\
&\theta_{q,n}=\arctan(y_{q,n}/x_{q,n}).
\end{aligned}\right. 
\end{equation}
The measurement noise $\tilde{\textbf{w}}_{q,n}$ is assumed to be AWGN, which is independent of $h(\bm{\xi}_{q,n})$ with zero means and the covariance matrix
\begin{equation}\label{Qm}
\bm{\Psi}_{q,n}=\text{diag}(\sigma^2_{d_{q,n}},\sigma^2_{v_{q,n}},\sigma^2_{\theta_{q,n}}),
\end{equation} 
where $\sigma^2_{d_{q,n}},\sigma^2_{v_{q,n}},\sigma^2_{\theta_{q,n}}$ are CRBs of the range, velocity, and AoA estimations. Similar to the definition (\ref{crb}), the CRB of velocity can be expressed as $\text{CRB}(v_{q,n}) = \beta_3/p_{q,n}|\varsigma_{q,n}|^2$. Since the dwell time allocation is not considered in this paper, the effective time duration $T_e$ is incorporated into coefficient $\beta_3$ as a system parameter. We see that the measurement covariance matrix $\bm{\Psi}_{q,n}$ depends on the power and bandwidth resources to be allocated.

\textit{PCRB:} Given a measurement $\textbf{y}_{q,n}$ w.r.t. a state $\bm{\xi}_{q,n}$, based on Bayes' theorem, the joint probability density function (PDF) of $\bm{\xi}_{q,n}$ and $\textbf{y}_{q,n}$ can be expressed as \cite{BookEstimationTheory}
\begin{equation}\label{PCRB1}
p(\bm{\xi}_{q,n},\textbf{y}_{q,n})=p(\textbf{y}_{q,n}|\bm{\xi}_{q,n})p(\bm{\xi}_{q,n}),
\end{equation} 
where $p(\textbf{y}_{q,n}|\bm{\xi}_{q,n})$ is the conditional PDF of $\textbf{y}_{q,n}$ with given $\bm{\xi}_{q,n}$, and $p(\bm{\xi}_{q,n})$ is the prior PDF of $\bm{\xi}_{q,n}$. Denote $\textbf{J}(\bm{\xi}_{q,n})$ as the posterior Fisher information matrix (FIM), which can be expressed as \cite{PCRB1998} 
\begin{equation}\label{PCRB2}
\textbf{J}(\bm{\xi}_{q,n})=-\mathbb{E}_{\bm{\xi}_{q,n},\textbf{y}_{q,n}}\left[ \nabla_{\bm{\xi}_{q,n}}^{\bm{\xi}_{q,n}} \ln p(\bm{\xi}_{q,n},\textbf{y}_{q,n})\right] ,
\end{equation}
where $\mathbb{E}_{\bm{\xi}_{q,n},\textbf{r}_{q,n}}[\cdot]$ is the expectation operation in terms of the state and the measurement vectors. $\nabla_{\Theta}^{\Psi}=\nabla_{\Theta}\nabla_{\Psi}^T$ represents the second-order partial derivatives operation. According to (\ref{PCRB1}), (\ref{PCRB2}), the original FIM can be divided into two parts  
\begin{equation}
\textbf{J}(\bm{\xi}_{q,n})=\textbf{J}_P(\bm{\xi}_{q,n})+\textbf{J}_D(\bm{\xi}_{q,n}),
\end{equation}  
where $\textbf{J}_P(\bm{\xi}_{q,n})$ and $\textbf{J}_D(\bm{\xi}_{q,n})$ are the prior information FIM and data FIM respectively, which can be calculated by \cite{PCRB1998}
\begin{equation}\label{FIMPD}
\left\{
\begin{aligned}
\textbf{J}_P(\bm{\xi}_{q,n})&=-\mathbb{E}_{\bm{\xi}_{q,n}}\left[ \nabla_{\bm{\xi}_{q,n}}^{\bm{\xi}_{q,n}} \ln p(\bm{\xi}_{q,n})\right] \\
& =\textbf{D}^{22}_{n-1}-\textbf{D}^{21}_{n-1}(\textbf{J}(\bm{\xi}_{q,n-1})+\textbf{D}^{11}_{n-1})^{-1}\textbf{D}^{12}_{n-1}, \\
\textbf{J}_D(\bm{\xi}_{q,n})&=-\mathbb{E}_{\bm{\xi}_{q,n},\textbf{y}_{q,n}}\left[ \nabla_{\bm{\xi}_{q,n}}^{\bm{\xi}_{q,n}} \ln p(\textbf{y}_{q,n}|\bm{\xi}_{q,n})\right]\\
& =\mathbb{E}_{\bm{\xi}_{q,n}}\left[\textbf{H}_{q,n}^T\bm{\Psi}_{q,n}^{-1}\textbf{H}_{q,n}\right],  
\end{aligned}\right. 
\end{equation}
where $\textbf{H}_{q,n}$ is the Jacobian matrix of the measurement function $h(\bm{\xi}_{q,n})$ w.r.t. the target state $\bm{\xi}_{q,n}$, and we have
\begin{equation}
\left\{
\begin{aligned}
\textbf{D}^{11}_{n-1}&=\textbf{F}_\xi^T\bm{\Phi}_q^{-1}\textbf{F}_\xi,   \\
\textbf{D}^{12}_{n-1}&=\textbf{D}^{21}_{n-1}=\textbf{F}_\xi^T\bm{\Phi}_q^{-1}, \\
\textbf{D}^{22}_{n-1}&=\bm{\Phi}_q^{-1}.
\end{aligned}\right.
\end{equation} 
The detailed derivations are omitted due to the space limitation. We refer the reader to \cite{PCRB1998} for more details. Formula (\ref{FIMPD}) implies that the $\textbf{J}_P$ at the $n$-th epoch can be calculated recursively with the given FIM $\textbf{J}$ at the first epoch. In addition, $\textbf{J}_D$ depends on the targets dynamics, and the resources to be allocated. Since the expectation operation is difficult to tackle in practice, the FIM can be approximated by \cite{Yan2015,liu2020radar}    
\begin{equation}
\textbf{J}_D(\bm{\xi}_{q,n})=\textbf{H}_{q,n}^T\bm{\Psi}_{q,n}^{-1}\textbf{H}_{q,n}|_{\bm{\xi}_{q,n}=\bm{\hat{\xi}}_{q,n|n-1}},
\end{equation}
where $\bm{\hat{\xi}}_{q,n|n-1}$ represent the predicted state based on the state evolution model by using the $(n-1)$-th state information. Thus, the FIM for the $q$-th user at the $n$-th epoch is the function w.r.t. the power and bandwidth resources, which can be rewritten as  
\begin{equation}\label{JFIM}
\begin{aligned}
\textbf{J}_{\bm{\xi}_{q,n}}(p_{q,n},b_{q,n})&=(\bm{\Phi}_q+\textbf{F}_\xi \textbf{J}^{-1}_{\bm{\xi}_{q,n-1}}\textbf{F}_\xi^T)^{-1} \\
&+\hat{\textbf{H}}_{q,n}^T\hat{\bm{\Psi}}_{q,n}^{-1}(p_{q,n},b_{q,n})\hat{\textbf{H}}_{q,n}.
\end{aligned}
\end{equation}
Consequently, the predicted PCRB for the state $\bm{\xi}_{q,n}$ can be given by
\begin{equation}\label{PCRBex}
\text{PCRB}_{\bm{\xi}_{q,n}}=\textbf{J}^{-1}_{\bm{\xi}_{q,n}}(p_{q,n},b_{q,n}).
\end{equation}

\subsection{Problem Formulation}

Similar to the RA problem for localization, our goal is to optimize the tracking QoS at the constraint of the communication QoS demands. Here, the objective parameter $\rho_{q,n}$ is determined by tracking QoS, i.e.,
\begin{equation}\label{OBP}
	\rho_{q,n} = \text{trace}(\text{PCRB}_{\bm{\xi}_{q,n}}).
\end{equation}
The associated optimization problem based on \textit{comprehensiveness} criterion can be formulated as
\begin{equation}\label{Tracking}
	\begin{aligned}
		\mathop { \text{minimize} }\limits_{\textbf{p}_n, \textbf{b}_n} \kern 10pt & F(\textbf{p}_n,\textbf{b}_n) =\sum \limits_{q=1}^{M_s} \rho_{q,n}  \\
		\text{subject to} \kern 10pt & R_n(\textbf{p}_n,\textbf{b}_n) \ge \Gamma_{c,n},  \\
		&\textbf{1}_{M}^T\textbf{p}_n=P_\text{total}, \textbf{1}_{M}^T\textbf{b}_n=B_\text{total}, \\
		&P_\text{min} \le p_{i,n} \le P_\text{max}, i=1,\cdots,M,  \\
		&B_\text{min} \le b_{j,n} \le B_\text{max}, j=1,\cdots,M,  \\
		&\rho_{1,n}:\cdots:\rho_{M_s,n}=\gamma_1:\cdots:\gamma_{M_s}. 	
	\end{aligned}
\end{equation}
where $\textbf{p}_n$ and $\textbf{b}_n$ are the power and bandwidth allocation vectors at the $n$-th epoch, and $\Gamma_{c,n}$ represents the communication QoS threshold at the $n$-th epoch. In addition, we omit the corresponding \textit{max-min} problem based on \textit{fairness} criterion which has a similar form with (\ref{ProDetectFair}).

As similar to the approach of solving the RA problem for localization, the AO method is again leveraged to decouple the variables. The rest task is to prove the convexity of each sub-problem. We give the explicit expression of PCRB w. r. t. power and bandwidth variables in Appendix, showing that the objective parameter $\rho_{q,n}$ is a convex function w. r. t. any single variable $\textbf{p}_n$ or $\textbf{b}_n$. Then, we have already shown that the constraints are convex in Section \ref{RAL} except for the last nonlinear equality constraints. To tackle these non-convex constraints, linearization process of the nonlinear constraints can be implemented at the $n$-th epoch, i.e., calculating the Taylor approximation of $\textbf{J}^{-1}_{\bm{\xi}_{q,n}}(\textbf{p}_n,\textbf{b}_n)$ at $(\textbf{p}_{n-1},\textbf{b}_{n-1})$. Consequently, each of the sub-problems is convex by linearizing the last non-linear equality constraints.

 \begin{figure*}[!t]	
	\subfigure{
		\begin{minipage}[t]{0.3\linewidth}
			\centering
			\includegraphics[width=2.3in]{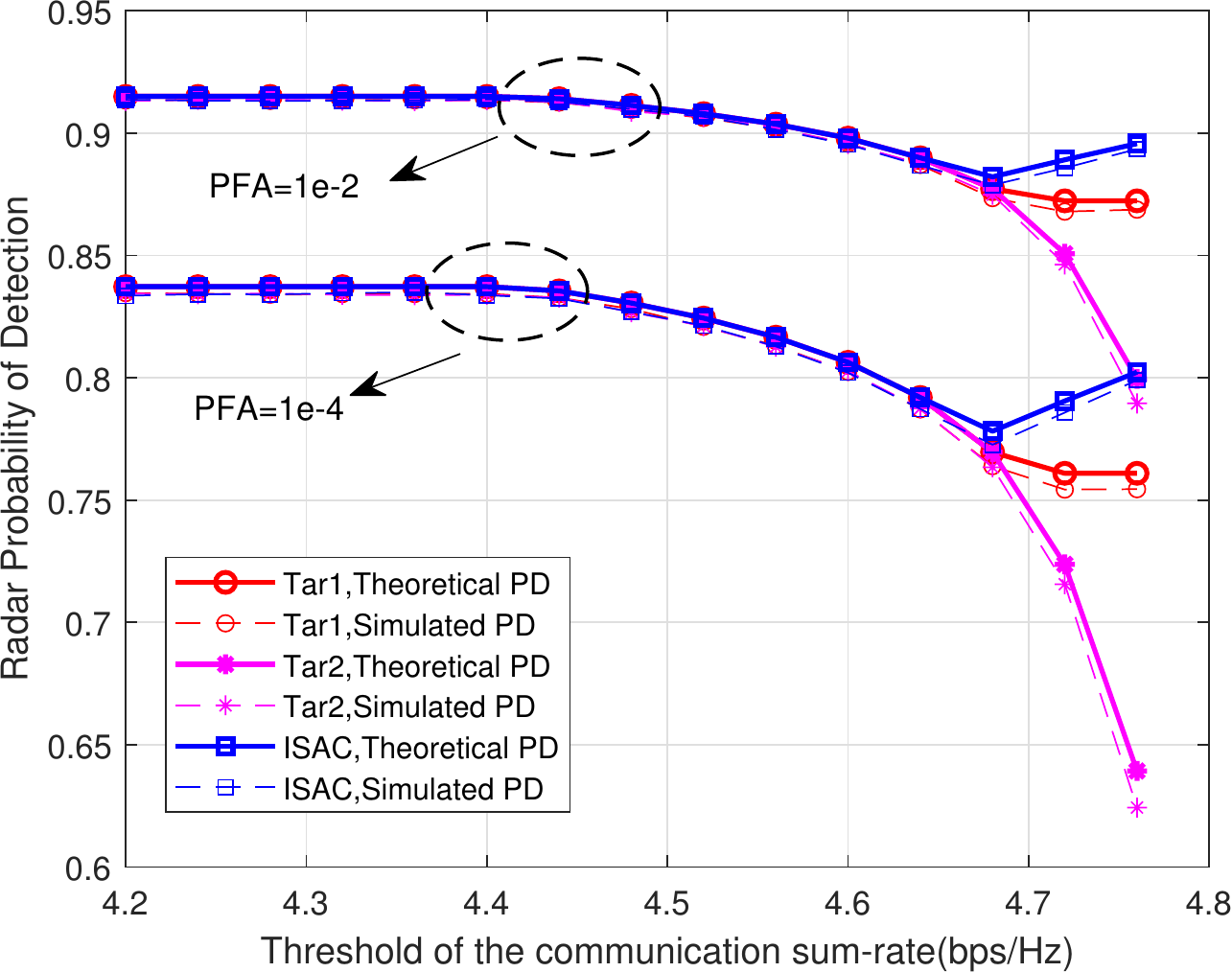}
			\caption{The sensing $P_\text{D}$ versus communication threshold in the \textit{fairness} scheme.}
			\label{PDMaxmin}
	\end{minipage}}
	\hspace{0.01\linewidth}	
	\subfigure{
		\begin{minipage}[t]{0.3\linewidth}
			\centering
			\includegraphics[width=2.3in]{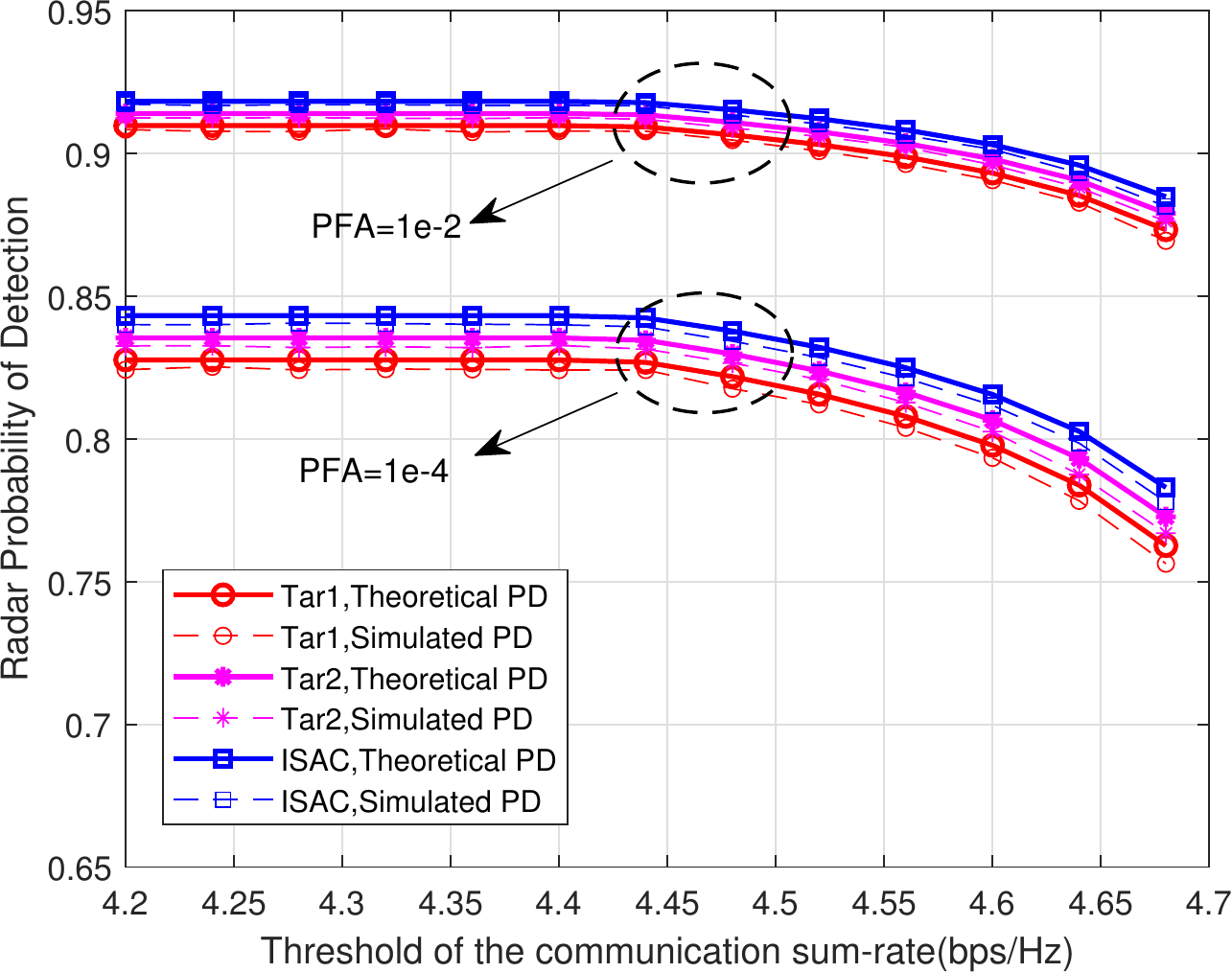}
			\caption{The sensing $P_\text{D}$ versus communication threshold in the \textit{comprehensiveness} scheme.}
			\label{PDSum}
	\end{minipage}}
	\hspace{0.01\linewidth} 
	\subfigure{
		\begin{minipage}[t]{0.3\linewidth}
			\centering
			\includegraphics[width=2.3in]{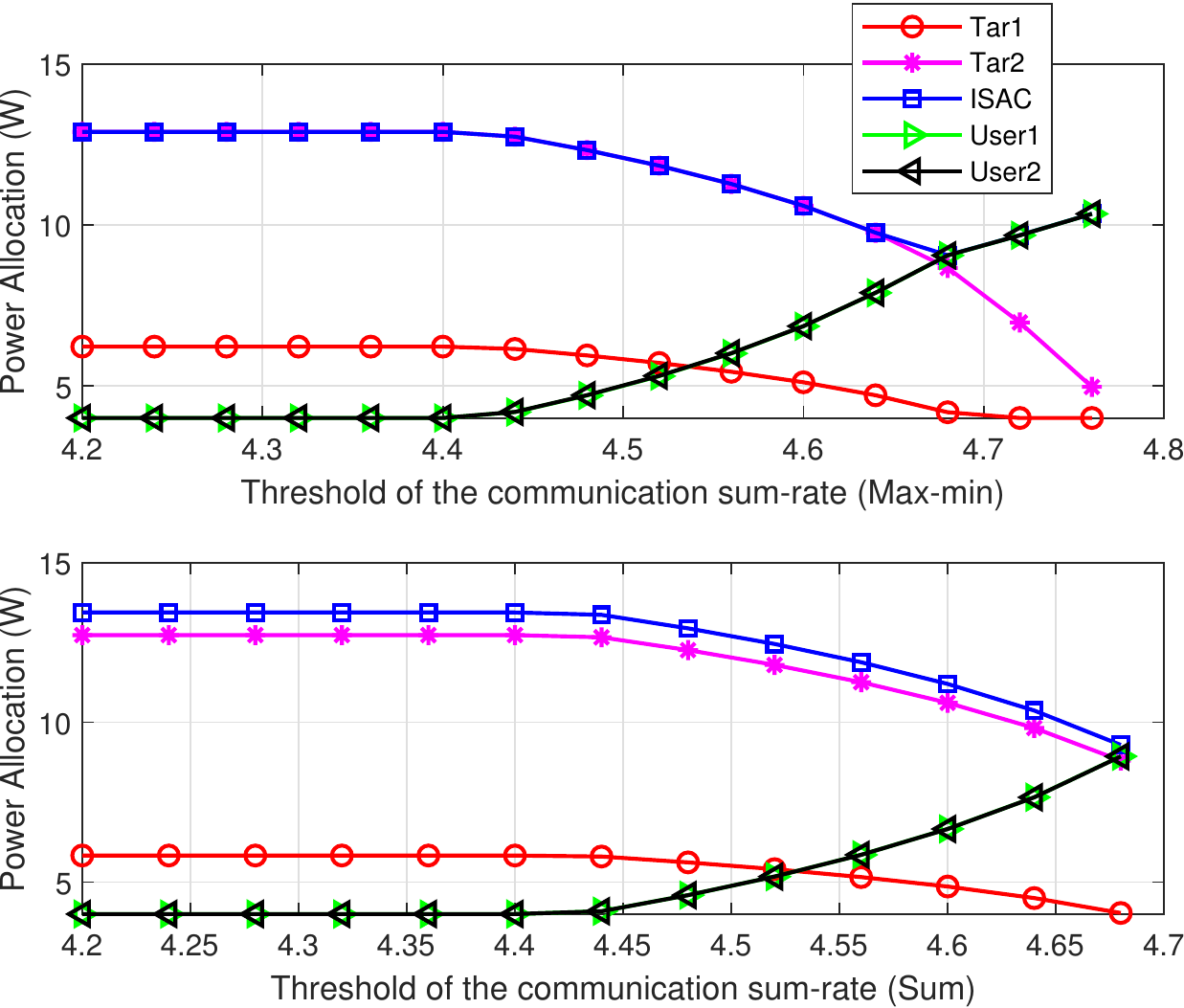}
			\caption{Power allocation results for the corresponding two schemes.}
			\label{PDPA}
	\end{minipage}}
\end{figure*}

Note that (\ref{OBP}) depends on the predicted state information, which is determined by the state evolution mode and the state at the last epoch. In order to achieve a good tracking performance, we establish a closed loop tracking strategy by adopting the extended Kalman filtering (EKF) scheme. Specifically, the predicted PCRBs in (\ref{OBP}) can be obtained after acquiring the target information at the current time. Then, the optimal power and bandwidth allocation for the next epoch can be attained through solving the optimization problem (\ref{Tracking}), resulting in optimal tracking accuracy. The procedure of EKF with RA is summarized as follows.

\textit{1) State Prediction:}
\begin{equation}
	\bm{\hat{\xi}}_{q,n|n-1}=\textbf{F}_\xi \bm{\xi}_{q,n-1}.
\end{equation}

\textit{2) MSE Matrix Prediction:}
\begin{equation}
	\textbf{M}_{q,n|n-1}=\textbf{F}_\xi \textbf{M}_{q,n-1} \textbf{F}_\xi^H + \bm{\Phi}_q.
\end{equation}
where the updated MSE matrix $\textbf{M}_{q,n-1}$ in the Kalman iteration equals to the predicted PCRB in (\ref{PCRBex}). (\textit{Theorem 1},\cite{liu2020radar})

\textit{3) Resource Allocation:} Allocate the resources $(\textbf{p}_n,\textbf{b}_n)$ by solving problem (\ref{ProLocal}) with tracking QoS ($\ref{OBP}$).  

\textit{4) Kalman Gain Calculation:}
\begin{equation}
	\textbf{K}_{q,n}=\textbf{M}_{q,n|n-1}\hat{\textbf{H}}_{q,n}^T\left( \hat{\bm{\Psi}}_{q,n} + \hat{\textbf{H}}_{q,n} \textbf{M}_{q,n|n-1} \hat{\textbf{H}}_{q,n}^T  \right)^{-1} .
\end{equation}

\textit{5) State Tracking:}
\begin{equation}
	\hat{\bm{\xi}}_{q,n}=\hat{\bm{\xi}}_{q,n|n-1}+\textbf{K}_{q,n}(\textbf{y}_{q,n}-h(\hat{\bm{\xi}}_{q,n|n-1})).
\end{equation}

\textit{6) MSE Matrix Update:}
\begin{equation}
	\textbf{M}_{q,n}=(\textbf{I} - \textbf{K}_{q,n} \hat{\textbf{H}}_{q,n})\textbf{M}_{q,n|n-1} .
\end{equation}

In summary, the BS is able to simultaneously provide S\&C services by iteratively performing prediction and tracking with a well-designed RA scheme.

\section{Simulation Results}\label{SR}

We evaluate the effectiveness of the proposed RA techniques by numerical simulation in this section. Unless otherwise specified, the BS is equipped with $N_t=N_r=32$ antennas, serving for multiple single-antenna users. The number of service objects is $M=5$ in total, including $Q=3$ sensing targets, $K=3$ communication users, and $L=1$ ISAC user. The frequency of carriers and the power spectral density are set as $f_c=30$ GHz and $\sigma_0 = -162$ dBm/Hz. The total amount of power resource is $P_\text{total}=40$ W, and that of bandwidth resource is $B_\text{total}=100$ MHz. In addition, the transmit power and bandwidth for each target/user are restricted within $[0.1P_\text{total},0.8P_\text{total}]$ and $[0.1B_\text{total},0.8B_\text{total}]$, respectively. The variance of path-loss for sensing targets is calculated according to the standard radar equation, i.e., $\hat{\sigma}^2_{\alpha_q}=\beta_\text{RCS}\lambda_c^2/((4\pi)^3d_q^4)$, where $\beta_\text{RCS}$ and $\lambda_c$ are the RCS of target and the wavelength of carrier, respectively. Moreover, the path-loss for communication user is calculated by $\tilde{\alpha}_{k}=32.4+20\log(d_k)+20\log(f_c)$ dB. The path-loss is mainly related to the distance between the BS and users when other parameters remain constant. Here, let $\textbf{d}=[d_1,\cdots,d_M]$ denote the distance vector.   

\begin{figure*}[htbp]
	
	\subfigure{
		\begin{minipage}[t]{0.3\linewidth}
			\centering
			\includegraphics[width=2.3in]{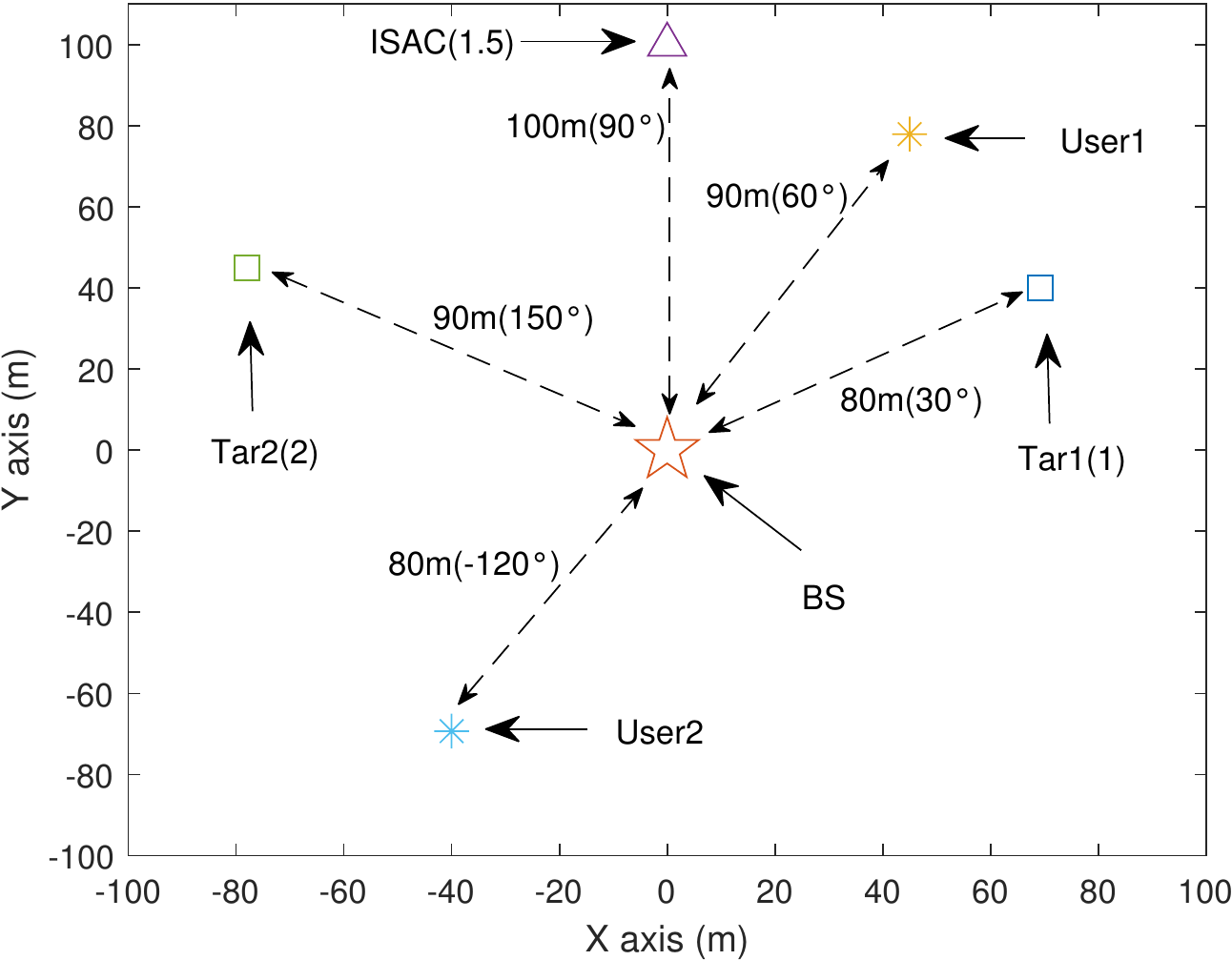}
			\caption{The geometric distribution of the targets and users in the localization task.}
			\label{diagram_Localization}
	\end{minipage}}
	\hspace{0.01\linewidth}	
	\subfigure{
		\begin{minipage}[t]{0.3\linewidth}
			\centering
			\includegraphics[width=2.3in]{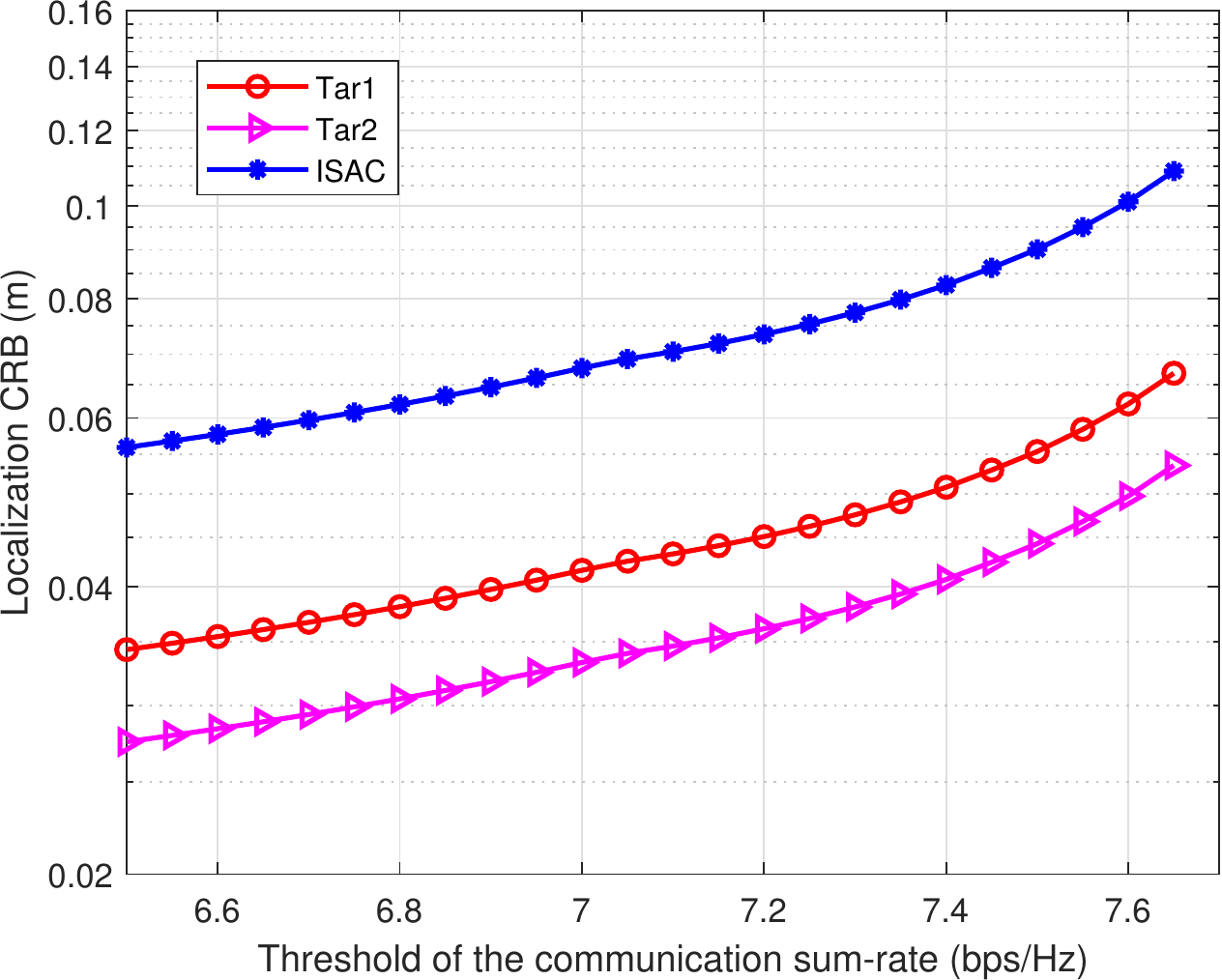}
			\caption{The localization CRBs for sensing targets versus communication threshold.}
			\label{Localization_CRB}
	\end{minipage}}
	\hspace{0.01\linewidth} 
	\subfigure{
		\begin{minipage}[t]{0.3\linewidth}
			\centering
			\includegraphics[width=2.3in]{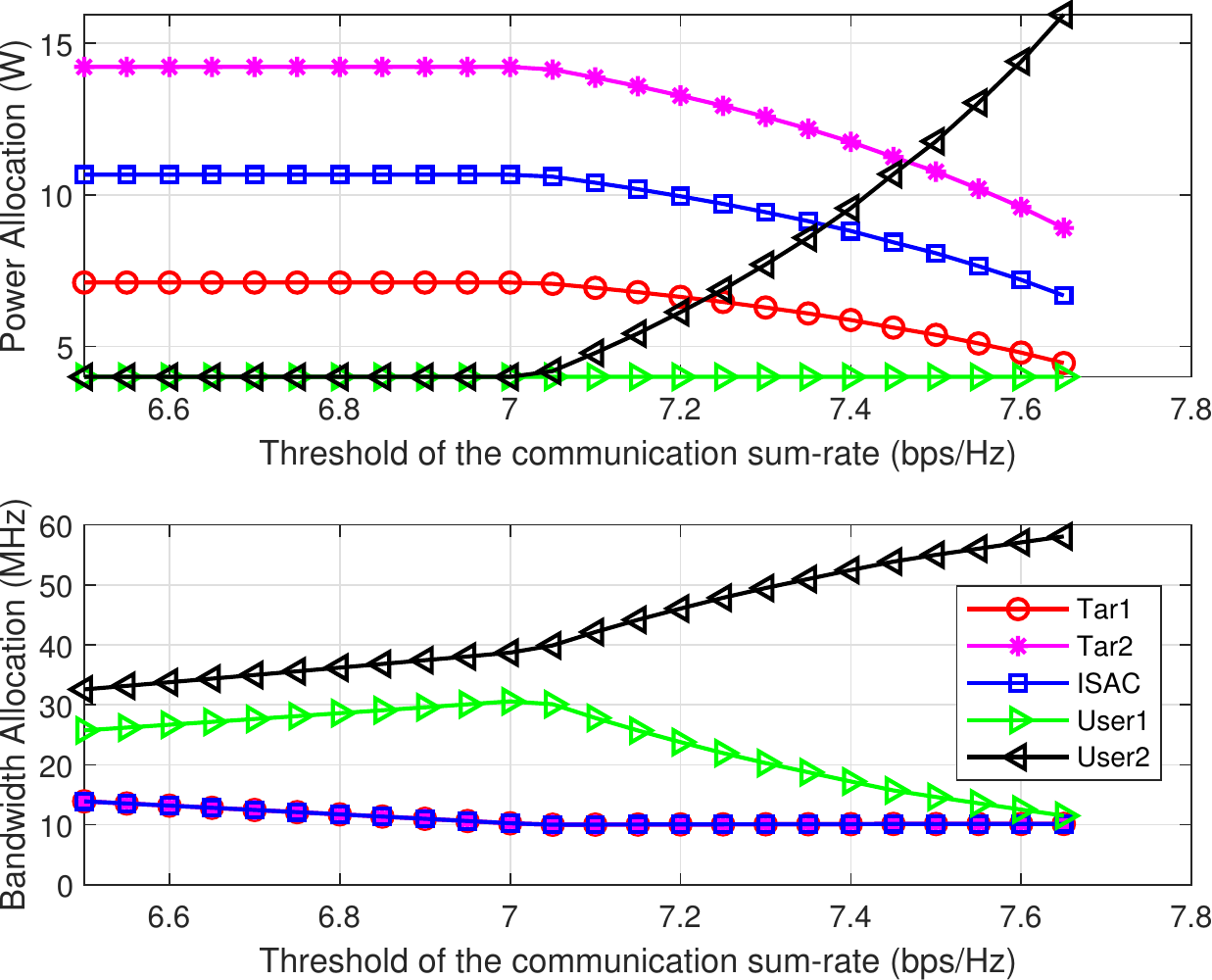}
			\caption{Power and bandwidth allocation results.}
			\label{Localization_RA}
	\end{minipage}}
\end{figure*}

\subsection{Target Detection}

In this subsection, we evaluate the performance trade-off between target detection and communication QoS. The distance parameter is set as $\textbf{d}=[100\text{m},120\text{m},120\text{m},110\text{m},100\text{m}]$. Unless otherwise specified, we assume that perfect beamforming is implemented, i.e., $\epsilon_q=\tilde{\epsilon}_q=1$ in the normalized sensing channel gain (\ref{CGN}). The theoretical $P_\text{D}$ is referred to calculating the $P_\text{D}$ by formula (\ref{PD}) for a given $P_\text{FA}$ and the power allocated. To show the effectiveness, the simulated $P_\text{D}$ is provided by constructing the signal detector to implement binary hypothesis test, where the results are obtained by $10^6$ Monte Carlo trials.     

Fig. \ref{PDMaxmin} shows the performance trade-off between sensing $P_\text{D}$ and communication threshold in \textit{fairness} scheme, and Fig. \ref{PDPA}(a) is the corresponding power allocation results. As shown in Fig. \ref{PDMaxmin}, the variation of $P_\text{D}$ can be divided into three stages upon the increasing communication sum-rate thresholds $\Gamma_c$. 1) At the first stage ($\Gamma_c<4.4$ bps/Hz), the $P_\text{D}$ of all sensing targets remain constant and equal at the region of small values of $\Gamma_c$, since the minimum power allocated to the communication users already meets the requirement. This can be verified in Fig. (\ref{PDPA}a) that the communication users are allocated the minimum power with $p=4$ W. 2) At the second stage ($\Gamma_c \in [4.4,4.68]$ bps/Hz), the $P_\text{D}$ for sensing targets still remains equal but decrease upon increasing $\Gamma_c$. It is because more power resource needs to be allocated for communication, leading to less power for sensing. 3) The bifurcation phenomenon occurs when $\Gamma_c > 4.68$ bps/Hz. Fig. \ref{PDPA}(a) shows that to achieve fairness, the power allocated to Tar1 is less than the others because it has the best channel quality ($d_1<d_2=d_3$). Thus, the power allocated to Tar1 first reaches the minimal value leading to the constant $P_\text{D}$, whereas the power for Tar2 continuously reduces. On the contrary, the power allocated to ISAC user increases to meet the communication sum-rate requirement, resulting in a increasing $P_\text{D}$ of sensing at the same time. In this case, the \textit{fairness} problem is no longer equivalent to the \textit{comprehensiveness} problem by setting the equal $\{\gamma_i\}_{i \in \mathcal{I}_s}$.        

Fig. \ref{PDSum} and Fig. \ref{PDPA}(b) provide the results for the \textit{comprehensiveness} scheme, where the fractional factors are set as $\bm{\gamma}=[1,0.95,0.9]$. In Fig. \ref{PDSum}, we can observe that the $P_\text{D}$ for sensing targets always changes in the preset proportion $\bm{\gamma}$. The ISAC user achieves the best detection QoS regardless of its poor channel quality, since we assign it a prior high importance level. Meanwhile, the power resource is also proportional allocated until the problem is infeasible. Finally, the low $P_\text{FA}$ results in a low $P_\text{D}$ as shown in both Fig. \ref{PDMaxmin} and Fig. \ref{PDSum} .

\subsection{Target Localization}

\begin{figure*}[htbp]	
	\subfigure{
		\begin{minipage}[t]{0.3\linewidth}
			\centering
			\includegraphics[width=2.3in]{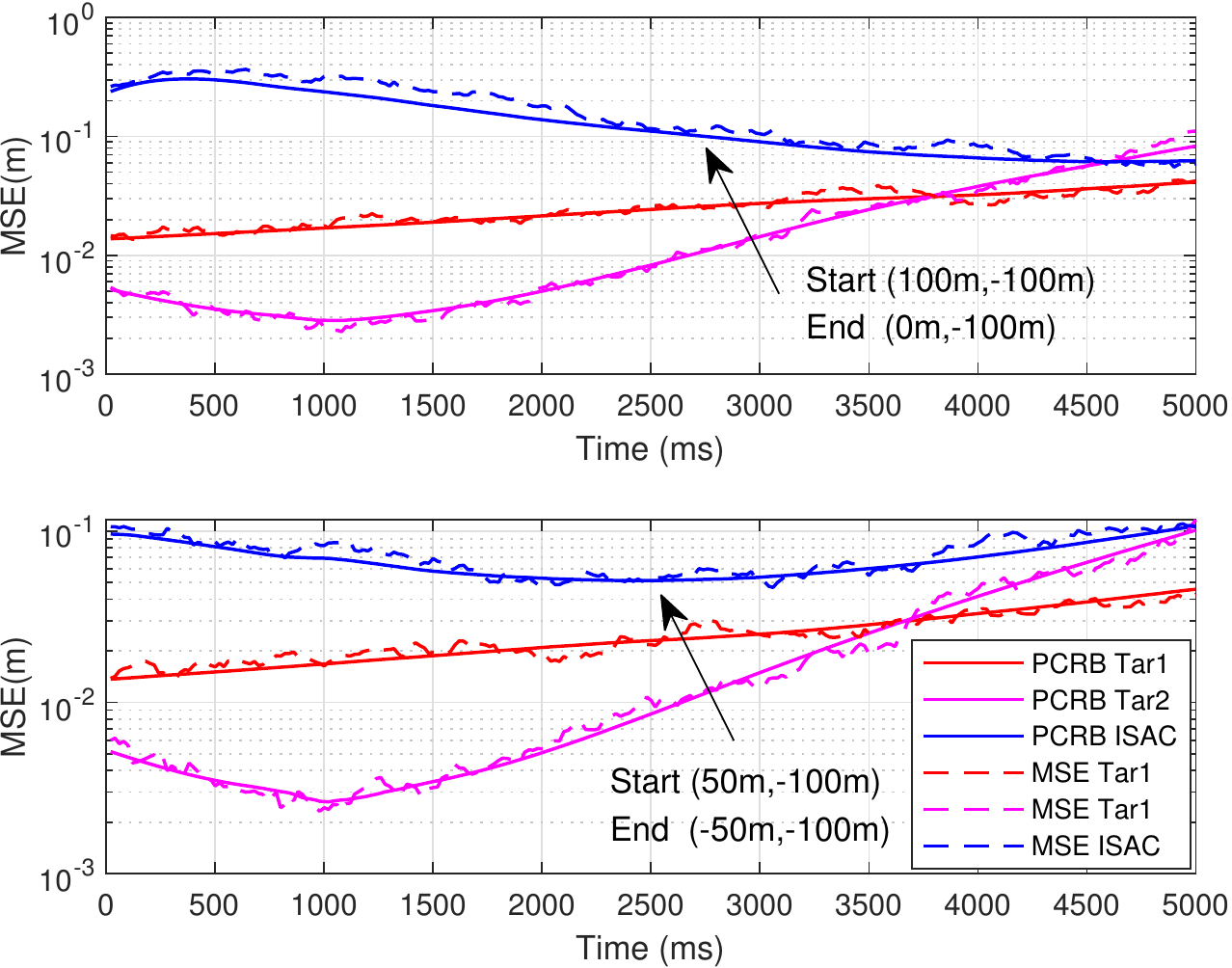}
			\caption{Distance tracking performance PCRB for the moving targets with $\Gamma_c=7.6$ bps/Hz.}
			\label{PCRB_tracking}
	\end{minipage}}
	\hspace{0.01\linewidth}	
	\subfigure{
		\begin{minipage}[t]{0.3\linewidth}
			\centering
			\includegraphics[width=2.3in]{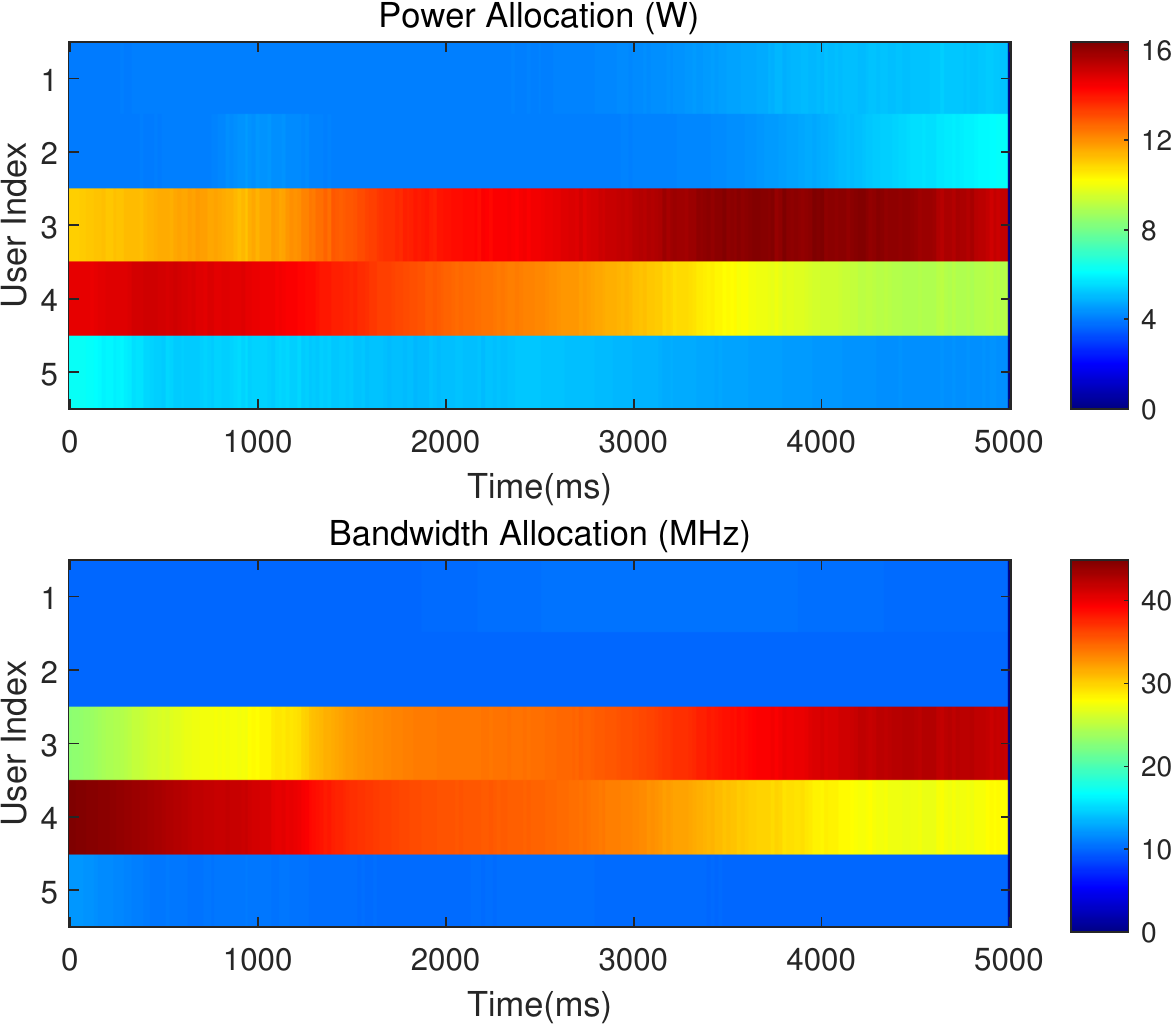}
			\caption{The RA results with the initial state $\bm{\xi_3'}$, $\Gamma_c=7.6$ bps/Hz.}
			\label{RA_tracking100}
	\end{minipage}}
	\hspace{0.01\linewidth} 
	\subfigure{
		\begin{minipage}[t]{0.3\linewidth}
			\centering
			\includegraphics[width=2.3in]{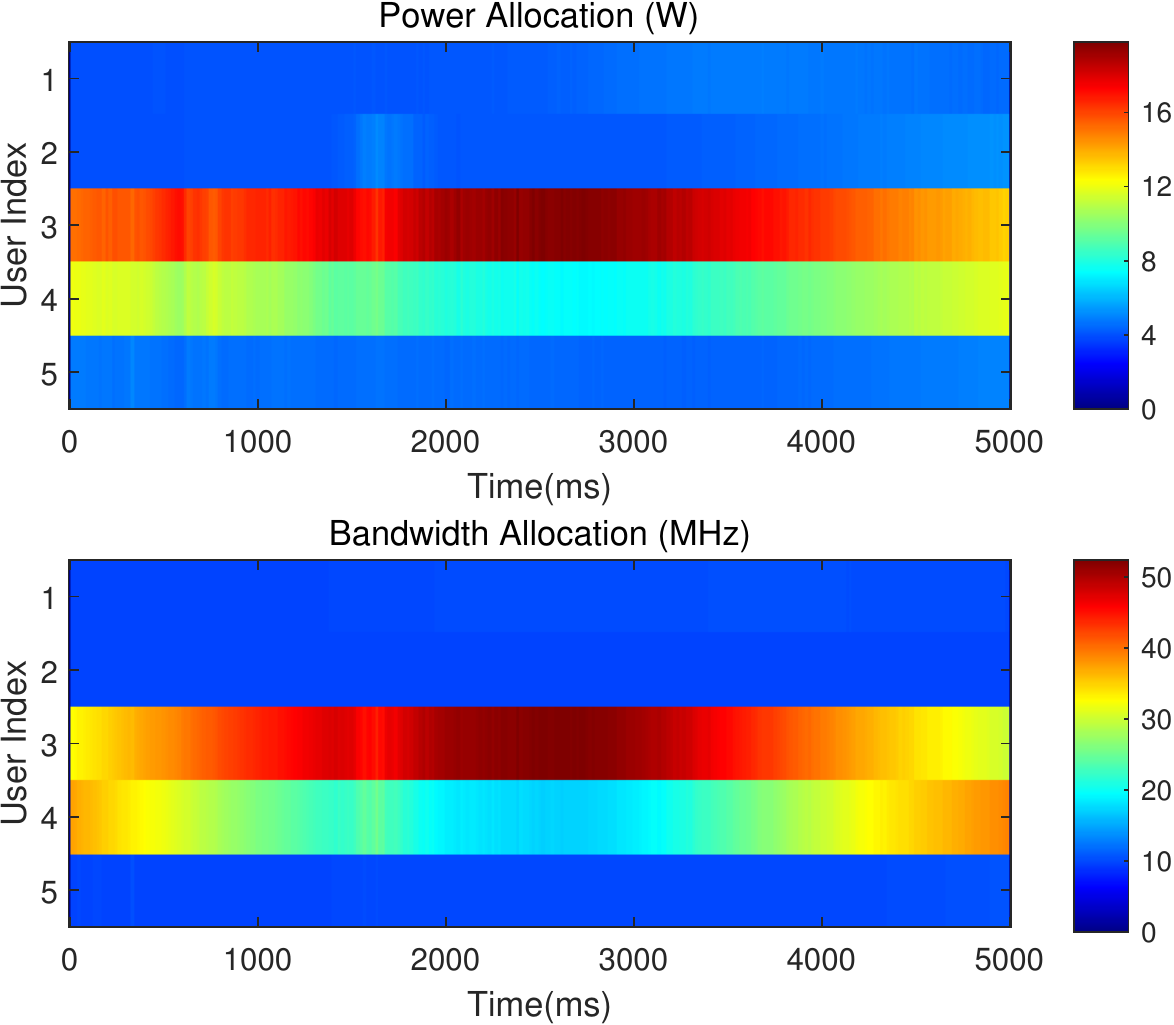}
			\caption{The RA results with the initial state $\bm{\xi_3}$, $\Gamma_c=7.6$ bps/Hz.}
			\label{RA_tracking50}			
	\end{minipage}}
	\setcounter{figure}{14}
\end{figure*}

The joint power and bandwidth allocation results for static targets localization are discussed in this subsection. In this group of simulations, we omit the \textit{max-min} problem to avoid repetition and focus on the RA for different sensing QoS. The distance is set as $d=[80\text{m},90\text{m},100\text{m},90\text{m},80\text{m}]$ and the AoA is $\bm{\theta}=[30^\circ,150^\circ,90^\circ,60^\circ,-120^\circ]$, where the geometric distribution of the targets and users are shown in Fig. \ref{diagram_Localization}. For localization, we consider the scenario that the BS steers its beam to the desired direction where there exists a potential target. Therefore, it is difficult to implement a perfect beamforming since we do not have accurate AoA information. For analysis convenience, we set $\epsilon_q= \tilde{\epsilon}_q=0.5, \forall q \in \mathcal{I}_s$ without loss of generality. In what follows, we degenerate the fractional constraint to $p_1b_1:p_qb_q=\gamma_1:\gamma_q$, to control the proportion of the resource to be allocated, where the fractional factors are set as $\bm{\gamma}=[1,1/2,2/3]$.
   
First of all, we should point out the necessity of the initial bandwidth selection process. The achievable sum-rate can reach 7.68 bps/Hz by calculating $R(\textbf{p}^{(0)},\textbf{b}^{(0)})$, where $\textbf{p}^{(0)}$ and $\textbf{b}^{(0)}$ are obtained by implementing the step 2 and 3 in Algorithm \ref{alg1}. However, if we allocate the equal bandwidth $B_\text{total}/M$ to each object (implement step 1 only), the problem will be infeasible when $\Gamma_c > 5.88$ bps/Hz. Fig. \ref{Localization_CRB} shows the performance trade-off between the QoS of localization and communication. We observe that the localization CRBs for all sensing targets increases upon the increasing communication threshold $\Gamma_c$ until the problem is infeasible. It is because more resources are used to meet high communication QoS, leading to relatively low localization accuracy. In particular, the Tar2 achieves the best localization QoS even though its path-loss ($d=90$m) is larger than Tar1's ($d=80$m) thanks to the proportional control constraint imposed, where the resources allocated to Tar2 is twice as large as that of Tar1. 

In Fig. \ref{Localization_RA}, the power and bandwidth results for the ISAC system are provided. It is shown that the power allocated to all users and targets remain constant at small thresholds with $\Gamma_c < 7.05$ bps/Hz. In this regime, the BS adjusts bandwidth allocation to meet communication QoS demands. By contrast, the power allocated to sensing targets start to reduce in a certain proportion when the threshold goes larger, since the bandwidth allocated to sensing targets reach the minimum value at $\Gamma_c = 7.05$ bps/Hz. As expected, both power and bandwidth allocated to User2 increase subsequently since it has the best channel gain among all the communication users.

\subsection{Target Tracking}

\begin{figure}[!t]
	\centering
	\includegraphics[width=3in]{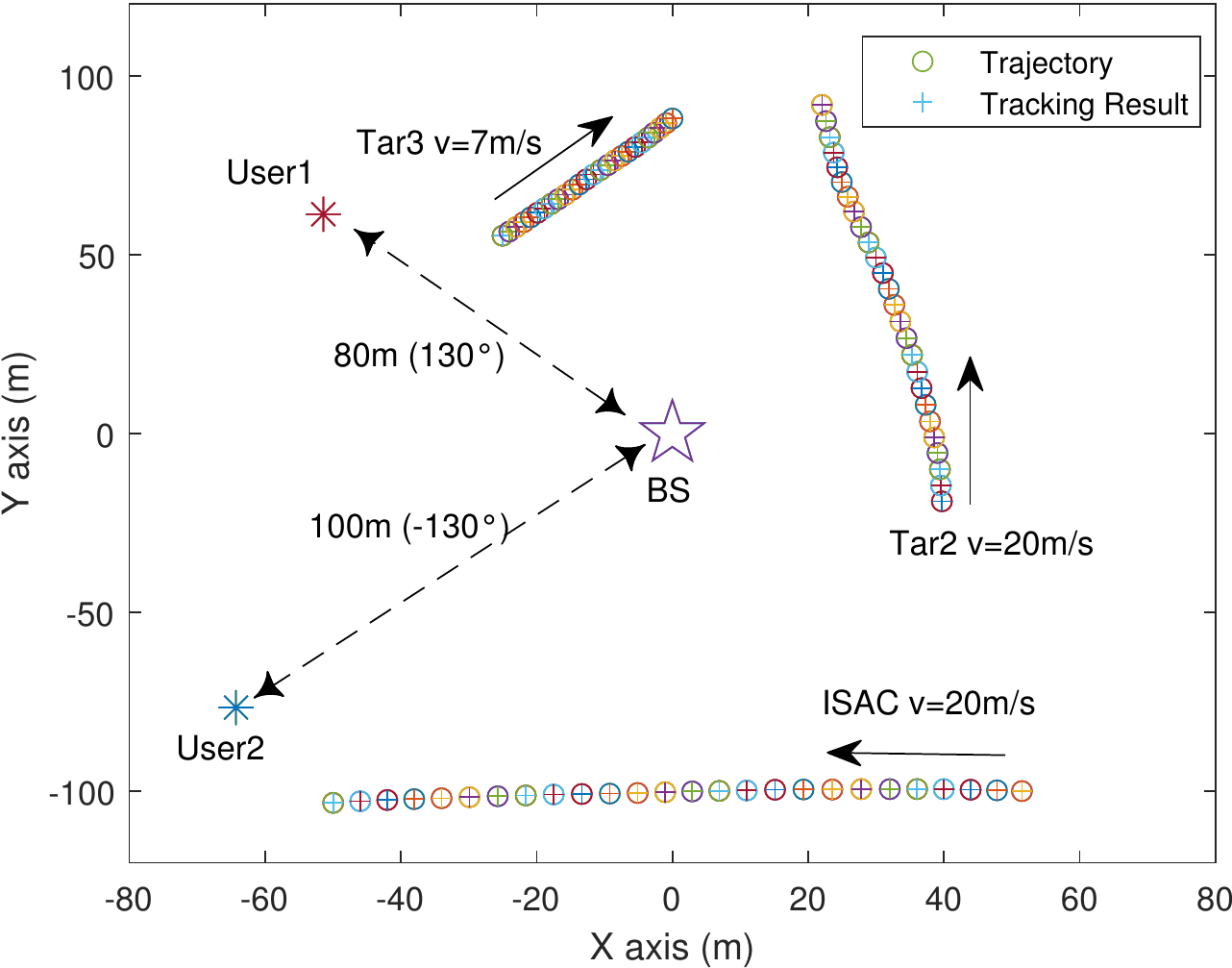}
	\caption{The geometric distribution of the targets and users as well as the trajectory and tracking results for the moving targets, $\Gamma_c=7.6$bps/Hz.}
	\label{diagram_tracking}
\end{figure}
In this subsection, we evaluate the performance of tracking and communication for the proposed RA scheme. Without loss of generality, the initial states of the targets to be sensed only are set as $\bm{\xi_1} = [-25\text{m}, 55\text{m}, 5\text{m/s}, 5\text{m/s}]$, $\bm{\xi_2} = [40\text{m}, -20\text{m}, 0\text{m/s}, 20\text{m/s}]$. The associated noise variances for state evolution are set as $\tilde{\sigma}_1=0.5, \tilde{\sigma}_2=2$ to describe different motion deviations. As for the ISAC user, the initial parameters are set as $\bm{\xi_3} = [50\text{m}, -100\text{m}, -20\text{m/s}, 0\text{m/s}]$ and $\tilde{\sigma}_3=1$. Moreover, the positions of the communication users are $80$m ($130^\circ$) and $100$m ($-130^\circ$). We use $\Delta T=0.02$ s as the block duration and the total period is $T=5$ s. The following results are obtained by averaging over $500$ Monte Carlo trials. 

From the above subsection, we can observe that the curves of sensing performance always vary in the preset proportion until the problem is infeasible in the \textit{comprehensiveness} scheme. In this subsection, we focus on the S\&C performance variation with the user's trajectory, hence the proportional constraints are temporarily omitted. Fig. \ref{diagram_tracking} shows the geometric distribution of the targets and users as well as the tracking results for the moving users in one trial. The trajectory and tracking results are picked every per $0.2$s. It can be seen that the proposed tracking scheme can achieve good tracking QoS even for the large state noise (Tar2).

Fig. \ref{PCRB_tracking} illustrates the PCRB of the targets' positions estimation versus time. A different start point of the ISAC user with $\bm{\xi_3'} = [100\text{m}, -100\text{m}]$ is chosen for performance comparison. For the initial state $\bm{\xi_3'}$, the ISAC user moves closer to the BS with the time, namely, the distance reduces from $141$m to $100$m. As expected, the MSE of the ISAC user decreases throughout the movement. The associated power and bandwidth allocation results are shown in Fig.\ref{RA_tracking100}. At the starting point, the ISAC user has the worst communication channel gain, so that the majority of the resources are allocated to User1 ($d=80$ m) to meet the communication QoS requirement. As the ISAC user moves closer to the BS, more resources are allocated to the ISAC user to simultaneously improve S\&C QoS. 

By contrast, for the initial state $\bm{\xi_3}$, the ISAC user starts from one side of the BS, then passes in front of the BS to its other side. Therefore, we can see that the MSE decreases initially and then increases over the time. Likewise, this phenomenon is caused by the change of the distance between the ISAC user and the BS, where the minimum MSE is achieved at the closest distance with $\theta=-90^\circ$. In Fig. \ref{RA_tracking50}, the trends of power and bandwidth allocation are consistent with the variation of MSE for the ISAC user. Additionally, note that the channel gain of the ISAC user is always superior to the other two communication users. Hence the resources are prioritized to the ISAC user all the time.         

\begin{figure}[!t]
	\centering
	\includegraphics[width=3in]{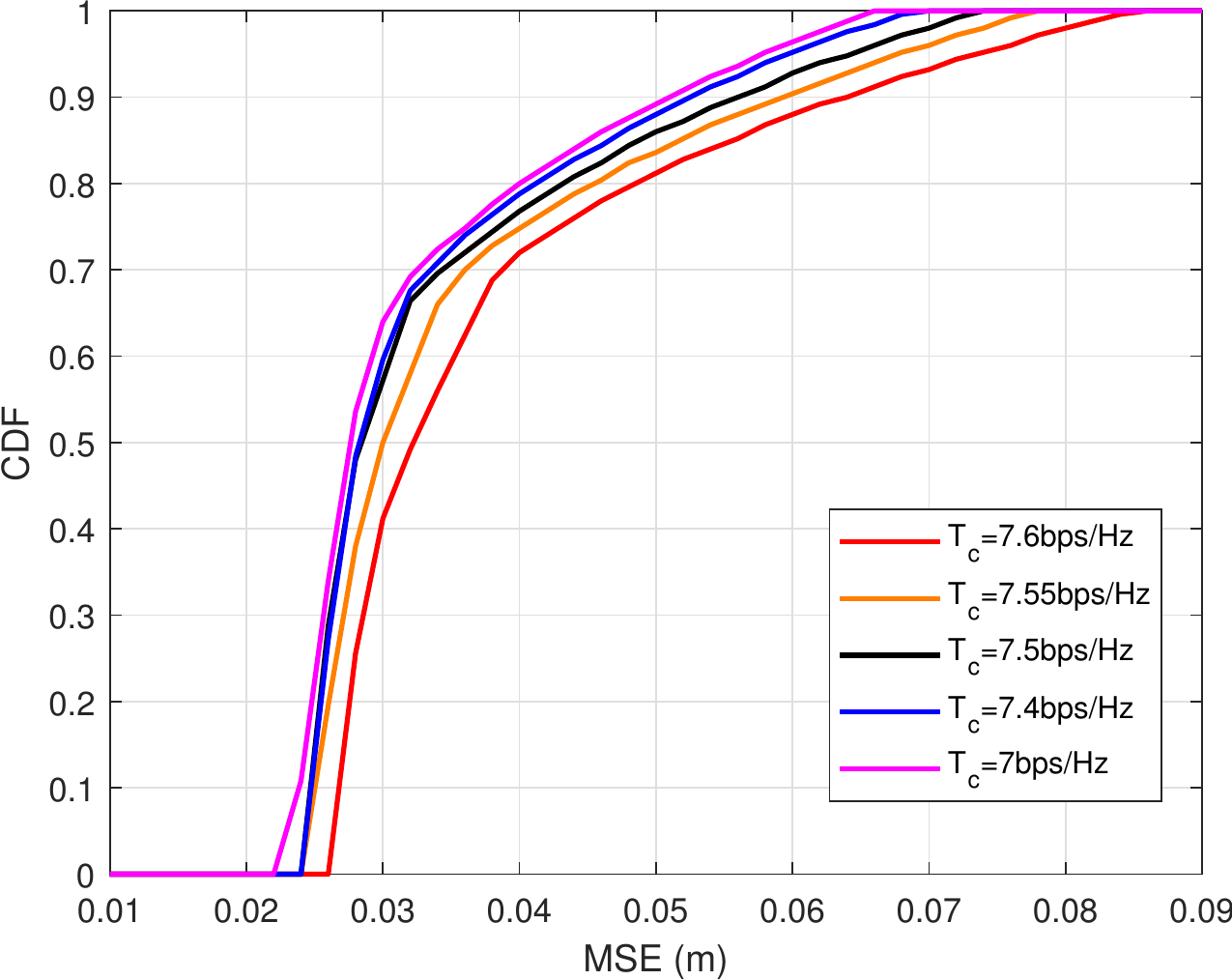}
	\caption{CDF of the MSE for all the sensing targets with different communication QoS requirements.}
	\label{CDF}
\end{figure}

Fig. \ref{CDF} demonstrates the performance trade-off between sensing and communication services in tracking. We provide the empirical cumulative distribution functions (CDF) in terms of the MSE for all the sensing targets. It shows that low communication QoS requirements lead to a high probability to achieve smaller MSE, thereby the better sensing QoS. We can observe that all of the curves do not start from the original point. In other words, the achievable MSEs of all sensing targets are always larger than a fixed value. It is because when the communication threshold $\Gamma_c$ is less than a certain value, the sensing QoS reaches its maximum since the resources allocated to communication users reach the minimum values, which implies that the communication QoS constraint is inactive.

\section{Conclusion}\label{Conclusion}
In this article, we have proposed a unified framework for integrated sensing and communication (ISAC) resource allocation in the upcoming 6G perceptive network. To measure the performance of sensing services, we have proposed the concept of sensing quality of service (QoS) for various applications. Specifically, the probability of detection, the Cr\'amer-Rao bound (CRB), and the posterior CRB (PCRB) are defined as the sensing QoS for detection, localization, and tracking, respectively. To achieve a scalable trade-off between sensing and communication QoS, the resource allocation problems based on fairness and comprehensiveness criteria are formulated and solved effectively. Numerical simulations have been provided to validate the proposed framework, which have clearly shown the trade-off between sensing and communication performance by adopting the desired resource allocation scheme.

\begin{appendix}  
	\section{}\label{A} 
	This is to prove the convexity of objective function (\ref{OBP}) w.r.t. the variable $\textbf{p}$ or $\textbf{b}$. After we acquire the FIM at the ($n-1$)-th epoch, formula (\ref{JFIM}) can be expressed in the matrix form of
	\begin{equation}\label{jpb}
	\textbf{J}_{\bm{\xi}_{q,n}}(p_{q,n},b_{q,n})=\textbf{E}+\textbf{H}^T\bm{\Lambda}(p_q,b_q) \textbf{H},
	\end{equation}
	where  $\bm{\Lambda}=\text{diag}(p_qb_q|\varsigma_q|^2/\beta_1,p_q|\varsigma_q|^2/\beta_2,p_q|\varsigma_q|^2/\beta_3)$ according to (\ref{Qm}), and $\textbf{E}=(\bm{\Phi}_q+\textbf{F}_\xi \textbf{J}^{-1}_{\bm{\xi}_{q,n-1}}\textbf{F}_\xi^T)^{-1}$ is a constant matrix. Here, we temporarily drop off the subscript $n$ for notation convenience. For a fixed $b_q$, we have
	\begin{equation}\label{50}
	\begin{aligned}
	\rho_q&=\text{trace}\left[ (\textbf{E}+p_q\textbf{V})^{-1}\right]   \\
 	   	  &=\text{trace}\left[ \textbf{E}^{-\frac{1}{2}}(\textbf{I}+p_q\textbf{E}^{-\frac{1}{2}}\textbf{V}\textbf{E}^{-\frac{1}{2}})^{-1}\textbf{E}^{-\frac{1}{2}}\right] \\
	      &=\text{trace}\left[ \textbf{E}^{-\frac{1}{2}}\textbf{U}(\textbf{I}+p_q\bm{\Xi})^{-1}\textbf{U}^T\textbf{E}^{-\frac{1}{2}}\right] \\
          &=\text{trace}\left[ \textbf{U}^T\textbf{E}^{-\frac{1}{2}}\textbf{E}^{-\frac{1}{2}}\textbf{U}(\textbf{I}+p_q\bm{\Xi})^{-1}\right] \\
          &=\textstyle\sum_{i=1}^{4}(\textbf{U}^T\textbf{E}^{-\frac{1}{2}}\textbf{E}^{-\frac{1}{2}}\textbf{U})_{ii}(1+p_q\Xi_{ii})^{-1}\\
          &=\textstyle\sum_{i=1}^{4}(a_{i,q}+b_{i,q}p_q)^{-1},
	\end{aligned}
	\end{equation}
	where $\textbf{V}=\textbf{H}^T\bm{\Lambda}\textbf{H}/p_q$ is the remaining matrix after extracting the common factor $p_q$. The eigenvalue decomposition with $\textbf{E}^{-1/2}\textbf{V}\textbf{E}^{-1/2}=\textbf{U}\bm{\Xi}\textbf{U}^T$ is implemented in (\ref{50}), where $\bm{\Xi}$ is a diagonal matrix with the elements of eigenvalues and the columns of $\textbf{U}$ are the corresponding eigenvectors. Here, the coefficients $a_{i,q}=1/(\textbf{U}^T\textbf{E}^{-1/2}\textbf{E}^{-1/2}\textbf{U})_{ii}$ and $b_{i,q}=\Xi_{ii}/a_{i,q}$ for $i=1,\cdots,4$ are all positive values thanks to the symmetric positive definite property of $\textbf{J}_D$ and $\textbf{J}_P$ \cite{liu2020radar,ZHW2020}, resulting in that $\rho_q$ is a convex function w.r.t. $p_q$. Accordingly, the \textit{comprehensiveness} objective function can be rewritten as 
	\begin{equation}
	F(\textbf{p}_n)= \sum \limits_{q=1}^{M_s}\sum_{i=1}^{4}(a_{i,q}+b_{i,q}p_q)^{-1},
	\end{equation}
	which is also convex since it is a linear combination of convex functions.    
	
	Similarly, for a fixed $p_q$, formula (\ref{jpb}) can be rewritten as 
	\begin{equation}\label{51}
	\begin{aligned}
	\textbf{J}_{\bm{\xi}_q}(b_{q})&=b_q\frac{p_q|\varsigma_q|^2}{\beta_1}\textbf{h}_1\textbf{h}_1^T+\sum_{i=2}^3\frac{p_q|\varsigma_q|^2}{\beta_i}\textbf{h}_i\textbf{h}_i^T+\textbf{E}\\
	&=b_q\textbf{V}_2+\textbf{E}_2.
	\end{aligned}
	\end{equation}	
	where $\textbf{h}_i$ is the $i$-th column of $\textbf{H}^T$, and 
	\begin{equation}
		\textbf{V}_2=\frac{p_q|\varsigma_q|^2}{\beta_1}\textbf{h}_1\textbf{h}_1^T,\kern 2pt \textbf{E}_2=\sum_{i=2}^3\frac{p_q|\varsigma_q|^2}{\beta_i}\textbf{h}_i\textbf{h}_i^T+\textbf{E}.
	\end{equation}
	It can be seen that (\ref{51}) has the similar form to (\ref{50}), hence its convexity can be immediately proven with the same process.      
		
\end{appendix}

\bibliographystyle{IEEEtran}
\bibliography{IEEEabrv,Resource_Allocation_R3}

\begin{thebibliography}{10}
\providecommand{\url}[1]{#1}
\csname url@samestyle\endcsname
\providecommand{\newblock}{\relax}
\providecommand{\bibinfo}[2]{#2}
\providecommand{\BIBentrySTDinterwordspacing}{\spaceskip=0pt\relax}
\providecommand{\BIBentryALTinterwordstretchfactor}{4}
\providecommand{\BIBentryALTinterwordspacing}{\spaceskip=\fontdimen2\font plus
\BIBentryALTinterwordstretchfactor\fontdimen3\font minus
  \fontdimen4\font\relax}
\providecommand{\BIBforeignlanguage}[2]{{%
\expandafter\ifx\csname l@#1\endcsname\relax
\typeout{** WARNING: IEEEtran.bst: No hyphenation pattern has been}%
\typeout{** loaded for the language `#1'. Using the pattern for}%
\typeout{** the default language instead.}%
\else
\language=\csname l@#1\endcsname
\fi
#2}}
\providecommand{\BIBdecl}{\relax}
\BIBdecl

\bibitem{Overview2020}
W.~Saad, M.~Bennis, and M.~Chen, ``A vision of 6g wireless systems:
  Applications, trends, technologies, and open research problems,'' \emph{IEEE
  Network}, vol.~34, no.~3, pp. 134--142, Jun. 2020.

\bibitem{ZAdrew2021}
A.~Zhang, M.~L. Rahman, X.~Huang, Y.~J. Guo, S.~Chen, and R.~W. Heath,
  ``Perceptive mobile networks: Cellular networks with radio vision via joint
  communication and radar sensing,'' \emph{IEEE Vehicular Technology Magazine},
  vol.~16, no.~2, pp. 20--30, Dec. 2021.

\bibitem{ZL2019}
L.~Zheng, M.~Lops, Y.~C. Eldar, and X.~Wang, ``Radar and communication
  coexistence: An overview: A review of recent methods,'' \emph{IEEE Signal
  Processing Magazine}, vol.~36, no.~5, pp. 85--99, Sept. 2019.

\bibitem{CC2020}
Z.~Feng, Z.~Fang, Z.~Wei, X.~Chen, Z.~Quan, and D.~Ji, ``Joint radar and
  communication: A survey,'' \emph{China Communications}, vol.~17, no.~1, pp.
  1--27, Jan. 2020.

\bibitem{DW2017}
A.~R. Chiriyath, B.~Paul, and D.~W. Bliss, ``Radar-communications convergence:
  Coexistence, cooperation, and co-design,'' \emph{IEEE Transactions on
  Cognitive Communications and Networking}, vol.~3, no.~1, pp. 1--12, Feb.
  2017.

\bibitem{LHzo2020}
F.~Liu, C.~Masouros, A.~P. Petropulu, H.~Griffiths, and L.~Hanzo, ``Joint radar
  and communication design: Applications, state-of-the-art, and the road
  ahead,'' \emph{IEEE Transactions on Communications}, vol.~68, no.~6, pp.
  3834--3862, Feb. 2020.

\bibitem{ComprehensiveS2021}
N.~C. Luong, X.~Lu, D.~T. Hoang, D.~Niyato, and D.~I. Kim, ``Radio resource
  management in joint radar and communication: A comprehensive survey,''
  \emph{IEEE Communications Surveys \& Tutorials}, vol.~23, no.~2, pp.
  780--814, Apr. 2021.

\bibitem{liu2021integrated}
F.~Liu, Y.~Cui, C.~Masouros, J.~Xu, T.~X. Han, Y.~C. Eldar, and S.~Buzzi,
  ``Integrated sensing and communications: Towards dual-functional wireless
  networks for 6g and beyond,'' \emph{arXiv preprint arXiv:2108.07165}, 2021.

\bibitem{CuiNet}
Y.~Cui, F.~Liu, X.~Jing, and J.~Mu, ``Integrating sensing and communications
  for ubiquitous iot: Applications, trends, and challenges,'' \emph{IEEE
  Network}, vol.~35, no.~5, pp. 158--167, Spet. 2021.

\bibitem{Huawei2021}
O.~Li, J.~He, K.~Zeng, Z.~Yu, X.~Du, Y.~Liang, G.~Wang, Y.~Chen, P.~Zhu,
  W.~Tong, D.~Lister, and L.~Ibbotson, ``Integrated sensing and communication
  in 6g a prototype of high resolution thz sensing on portable device,'' in
  \emph{2021 Joint European Conference on Networks and Communications 6G Summit
  (EuCNC/6G Summit)}, 2021, pp. 544--549.

\bibitem{ZJ2021}
J.~A. Zhang, M.~L. Rahman, K.~Wu, X.~Huang, Y.~J. Guo, S.~Chen, and J.~Yuan,
  ``Enabling joint communication and radar sensing in mobile networks: A
  survey,'' \emph{IEEE Communications Surveys \& Tutorials}, pp. 1--1, 2021.

\bibitem{Automous2020}
D.~Ma, N.~Shlezinger, T.~Huang, Y.~Liu, and Y.~C. Eldar, ``Joint
  radar-communication strategies for autonomous vehicles: Combining two key
  automotive technologies,'' \emph{IEEE Signal Processing Magazine}, vol.~37,
  no.~4, pp. 85--97, Jun. 2020.

\bibitem{IoT2021}
H.~Hong, J.~Zhao, T.~Hong, and T.~Tang, ``Radar-communication integration for
  6g massive iot services,'' \emph{IEEE Internet of Things Journal}, pp. 1--1,
  2021.

\bibitem{BookDetectionTheory}
S.~M. Kay, \emph{Fundamentals of statistical signal processing: Detection
  theory}.\hskip 1em plus 0.5em minus 0.4em\relax Englewood Cliffs, NJ:
  Prentice-Hall, 1993.

\bibitem{BookEstimationTheory}
------, \emph{Fundamentals of statistical signal processing: Estimation
  theory}.\hskip 1em plus 0.5em minus 0.4em\relax Englewood Cliffs, NJ:
  Prentice-Hall, 1993.

\bibitem{PCRB1998}
P.~Tichavsky, C.~Muravchik, and A.~Nehorai, ``Posterior cramer-rao bounds for
  discrete-time nonlinear filtering,'' \emph{IEEE Transactions on Signal
  Processing}, vol.~46, no.~5, pp. 1386--1396, May 1998.

\bibitem{KDcom2003}
D.~Kivanc, G.~Li, and H.~Liu, ``Computationally efficient bandwidth allocation
  and power control for ofdma,'' \emph{IEEE Transactions on Wireless
  Communications}, vol.~2, no.~6, pp. 1150--1158, Nov. 2003.

\bibitem{Propor2005}
Z.~Shen, J.~Andrews, and B.~Evans, ``Adaptive resource allocation in multiuser
  ofdm systems with proportional rate constraints,'' \emph{IEEE Transactions on
  Wireless Communications}, vol.~4, no.~6, pp. 2726--2737, Nov. 2005.

\bibitem{LTERA2013}
P.~Phunchongharn, E.~Hossain, and D.~I. Kim, ``Resource allocation for
  device-to-device communications underlaying lte-advanced networks,''
  \emph{IEEE Wireless Communications}, vol.~20, no.~4, pp. 91--100, Sept. 2013.

\bibitem{GRA2017}
H.~Zhang, N.~Liu, X.~Chu, K.~Long, A.-H. Aghvami, and V.~C.~M. Leung, ``Network
  slicing based 5g and future mobile networks: Mobility, resource management,
  and challenges,'' \emph{IEEE Communications Magazine}, vol.~55, no.~8, pp.
  138--145, Aug. 2017.

\bibitem{NOMARA2018}
S.~M.~R. Islam, M.~Zeng, O.~A. Dobre, and K.-S. Kwak, ``Resource allocation for
  downlink noma systems: Key techniques and open issues,'' \emph{IEEE Wireless
  Communications}, vol.~25, no.~2, pp. 40--47, Apr. 2018.

\bibitem{Yan2015}
J.~Yan, H.~Liu, B.~Jiu, B.~Chen, Z.~Liu, and Z.~Bao, ``Simultaneous multibeam
  resource allocation scheme for multiple target tracking,'' \emph{IEEE
  Transactions on Signal Processing}, vol.~63, no.~12, pp. 3110--3122, Mar.
  2015.

\bibitem{Yan2017cooperative}
J.~Yan, W.~Pu, H.~Liu, S.~Zhou, and Z.~Bao, ``Cooperative target assignment and
  dwell allocation for multiple target tracking in phased array radar
  network,'' \emph{Signal Processing}, vol. 141, pp. 74--83, 2017.

\bibitem{ZHW2020}
H.~Zhang, B.~Zong, and J.~Xie, ``Power and bandwidth allocation for
  multi-target tracking in collocated mimo radar,'' \emph{IEEE Transactions on
  Vehicular Technology}, vol.~69, no.~9, pp. 9795--9806, Sept. 2020.

\bibitem{SPL2017}
B.~K. Chalise, M.~G. Amin, and B.~Himed, ``Performance tradeoff in a unified
  passive radar and communications system,'' \emph{IEEE Signal Processing
  Letters}, vol.~24, no.~9, pp. 1275--1279, Jun. 2017.

\bibitem{LY2017}
Y.~Liu, G.~Liao, J.~Xu, Z.~Yang, and Y.~Zhang, ``Adaptive ofdm integrated radar
  and communications waveform design based on information theory,'' \emph{IEEE
  Communications Letters}, vol.~21, no.~10, pp. 2174--2177, July 2017.

\bibitem{XYF2018}
Y.~Xiong, N.~Wu, Y.~Shen, and M.~Z. Win, ``Cooperative network synchronization:
  Asymptotic analysis,'' \emph{IEEE Transactions on Signal Processing},
  vol.~66, no.~3, pp. 757--772, Oct. 2018.

\bibitem{XYF2022}
------, ``Cooperative localization in massive networks,'' \emph{IEEE
  Transactions on Information Theory}, vol.~68, no.~2, pp. 1237--1258, Nov.
  2022.

\bibitem{RadarSP}
M.~A. Richards, \emph{Fundamentals of radar signal processing, second
  edition}.\hskip 1em plus 0.5em minus 0.4em\relax McGraw-Hill Education, 2014.

\bibitem{FE2006}
E.~Fishler, A.~Haimovich, R.~Blum, L.~Cimini, D.~Chizhik, and R.~Valenzuela,
  ``Spatial diversity in radars—models and detection performance,''
  \emph{IEEE Transactions on Signal Processing}, vol.~54, no.~3, pp. 823--838,
  Mar. 2006.

\bibitem{IB2006}
I.~Bekkerman and J.~Tabrikian, ``Target detection and localization using mimo
  radars and sonars,'' \emph{IEEE Transactions on Signal Processing}, vol.~54,
  no.~10, pp. 3873--3883, Oct. 2006.

\bibitem{CVX}
\BIBentryALTinterwordspacing
S.~B. M.~Grant. (2020) Cvx: Matlab software for disciplined convex programming,
  version 2.2. [Online]. Available: \url{http://cvxr.com/cvx.}
\BIBentrySTDinterwordspacing

\bibitem{liu2021survey}
A.~Liu, Z.~Huang, M.~Li, Y.~Wan, W.~Li, T.~X. Han, C.~Liu, R.~Du, D.~K.~P. Tan,
  J.~Lu, Y.~Shen, F.~Colone, and K.~Chetty, ``A survey on fundamental limits of
  integrated sensing and communication,'' \emph{IEEE Communications Surveys \&
  Tutorials}, vol.~24, no.~2, pp. 994--1034, 2022.

\bibitem{MS1960}
M.~I. Skolnik, ``Theoretical accuracy of radar measurements,'' \emph{IRE
  Transactions on Aeronautical and Navigational Electronics}, no.~4, pp.
  123--129, 1960.

\bibitem{ngo2015massive}
H.~Q. Ngo, \emph{Massive MIMO: Fundamentals and system designs}.\hskip 1em plus
  0.5em minus 0.4em\relax Link{\"o}ping University Electronic Press, 2015, vol.
  1642.

\bibitem{liu2020radar}
F.~Liu, W.~Yuan, C.~Masouros, and J.~Yuan, ``Radar-assisted predictive
  beamforming for vehicular links: Communication served by sensing,''
  \emph{IEEE Transactions on Wireless Communications}, vol.~19, no.~11, pp.
  7704--7719, Nov. 2020.

\end{thebibliography}

\begin{IEEEbiography}[{\includegraphics[width=1in,height=1.25in,clip,keepaspectratio]{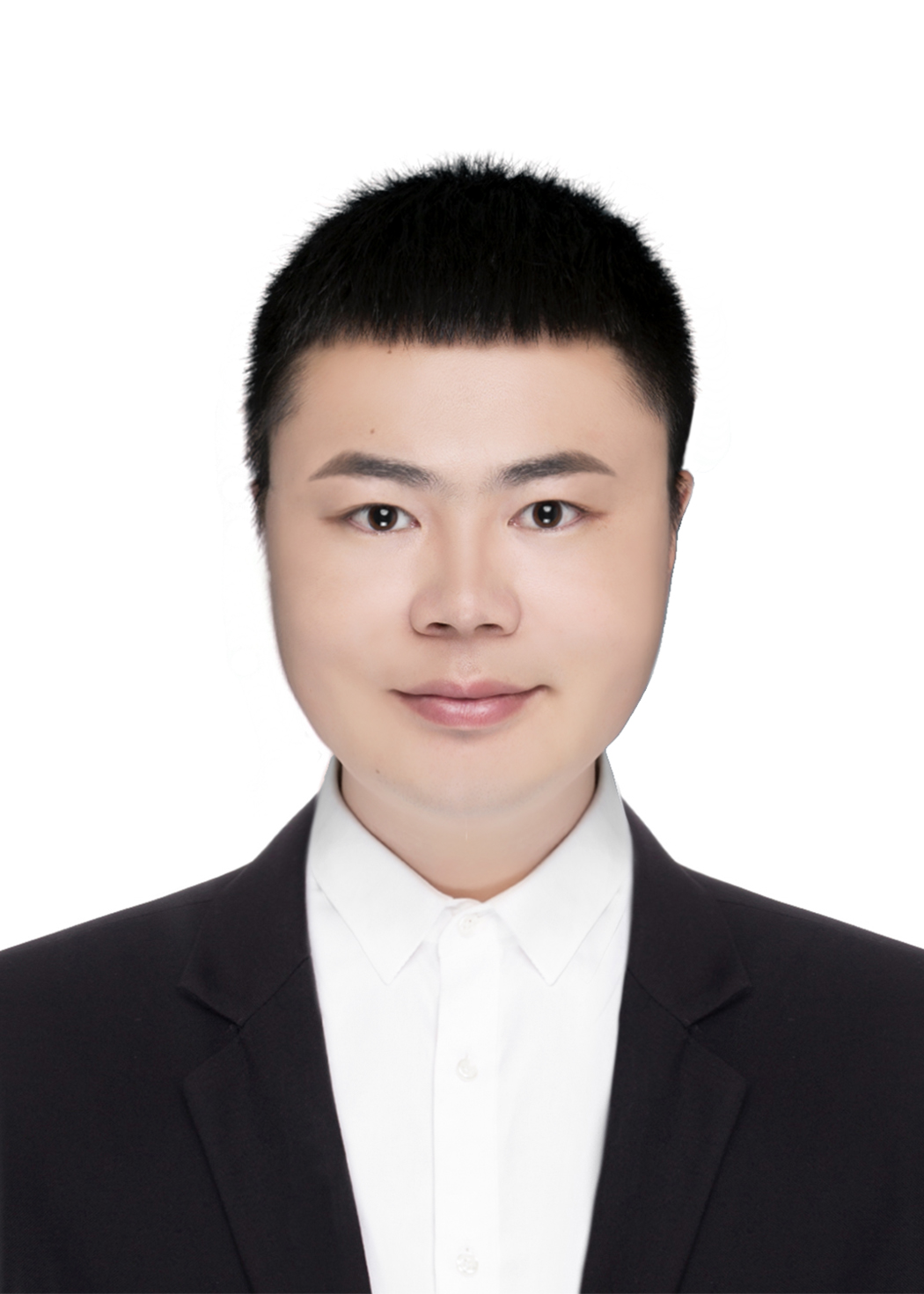}}]{Fuwang Dong} (Member, IEEE) is currently a post-doctoral of the Department of Electronic and Electrical Engineering, Southern University of Science and Technology (SUSTech). He received the Ph.D., the MSc., and the BSc. degrees from Harbin Engineering University (HEU), Harbin, China, in 2022, 2017, and 2014, respectively. His research interests include integrated sensing and communications (ISAC), array signal processing, hybrid precoding for millimeter wave communication systems, etc.
	
\end{IEEEbiography}

\begin{IEEEbiography}[{\includegraphics[width=1in,height=1.25in,clip,keepaspectratio]{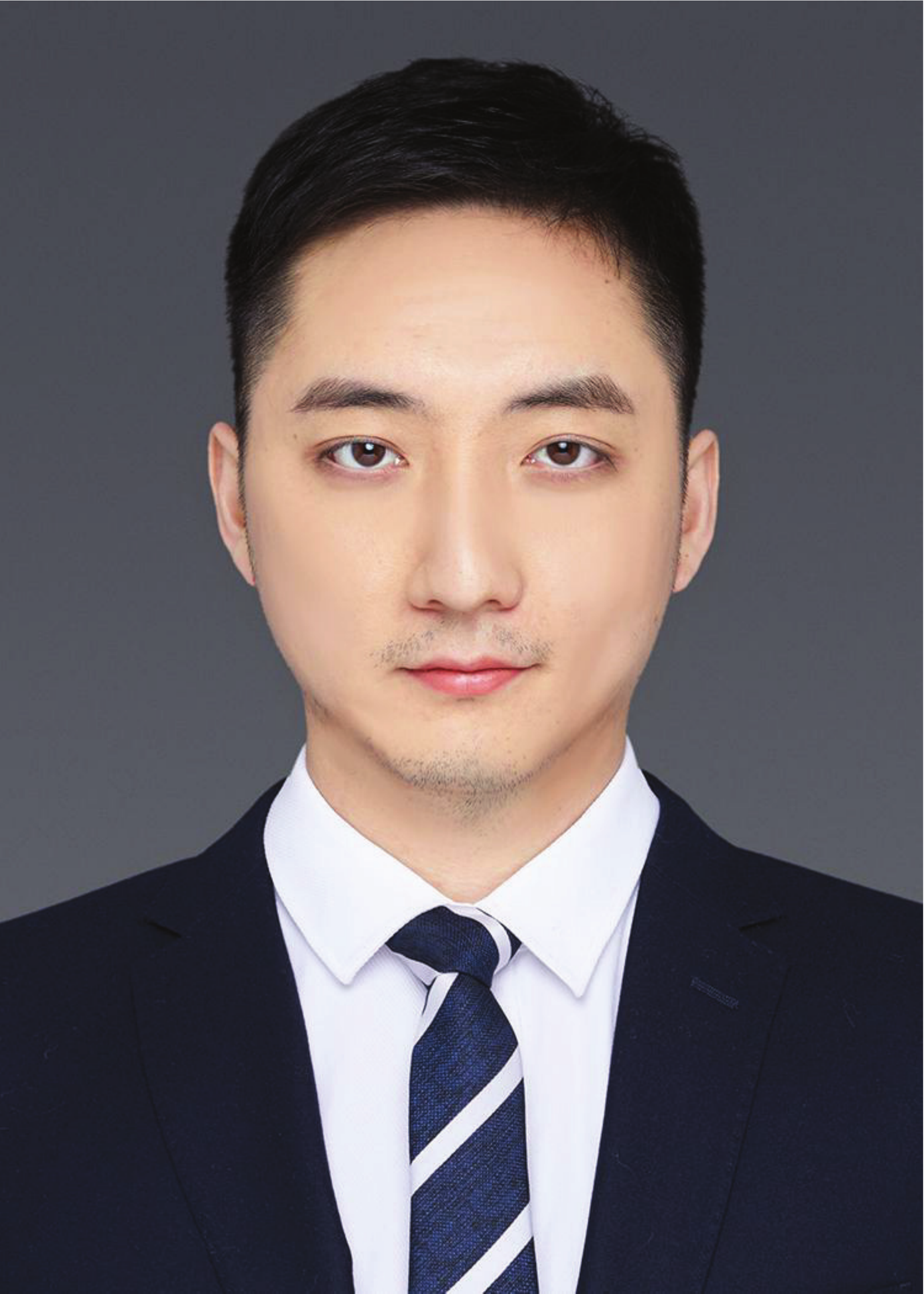}}]{Fan Liu} (Member, IEEE) is currently an Assistant Professor of the Department of Electronic and Electrical Engineering, Southern University of Science and Technology (SUSTech). He received the Ph.D. and the BEng. degrees from Beijing Institute of Technology (BIT), Beijing, China, in 2018 and 2013, respectively. He has previously held academic positions in the University College London (UCL), first as a Visiting Researcher from 2016 to 2018, and then as a Marie Curie Research Fellow from 2018 to 2020.
	
Dr. Fan Liu's research interests include the general area of signal processing and wireless communications, and in particular in the area of Integrated Sensing and Communications (ISAC). He has 10 publications selected as IEEE ComSoc Besting Readings in ISAC. He is the Founding Academic Chair of the IEEE ComSoc ISAC Emerging Technology Initiative (ISAC-ETI), an Associate Editor for the IEEE Communications Letters and the IEEE Open Journal of Signal Processing, a Lead Guest Editor of the IEEE Journal on Selected Areas in Communications, special issue on ``Integrated Sensing and Communication'', and a Guest Editor of the IEEE Wireless Communications, special issue on ``Integrated Sensing and Communications for 6G''. He was also an organizer and Co-Chair for numerous workshops, special sessions and tutorials in flagship IEEE/ACM conferences, including ICC, GLOBECOM, ICASSP, and MobiCom. He is the TPC Co-Chair of the 2nd and 3rd IEEE Joint Communication and Sensing Symposium (JC\&S), and will serve as a Track Co-Chair for the IEEE WCNC 2024. He is a Member of the IMT-2030 (6G) ISAC Task Group. Dr. Fan Liu was listed in the World's Top 2\% Scientists by Stanford University for citation impact in 2021 and 2022. He was also the recipient of the IEEE Signal Processing Society Young Author Best Paper Award of 2021, the Best Ph.D. Thesis Award of Chinese Institute of Electronics of 2019, the EU Marie Curie Individual Fellowship in 2018, and has been named as an Exemplary Reviewer for IEEE TWC/TCOM/COMML for 5 times.
\end{IEEEbiography}

\begin{IEEEbiography}[{\includegraphics[width=1in,height=1.25in,clip,keepaspectratio]{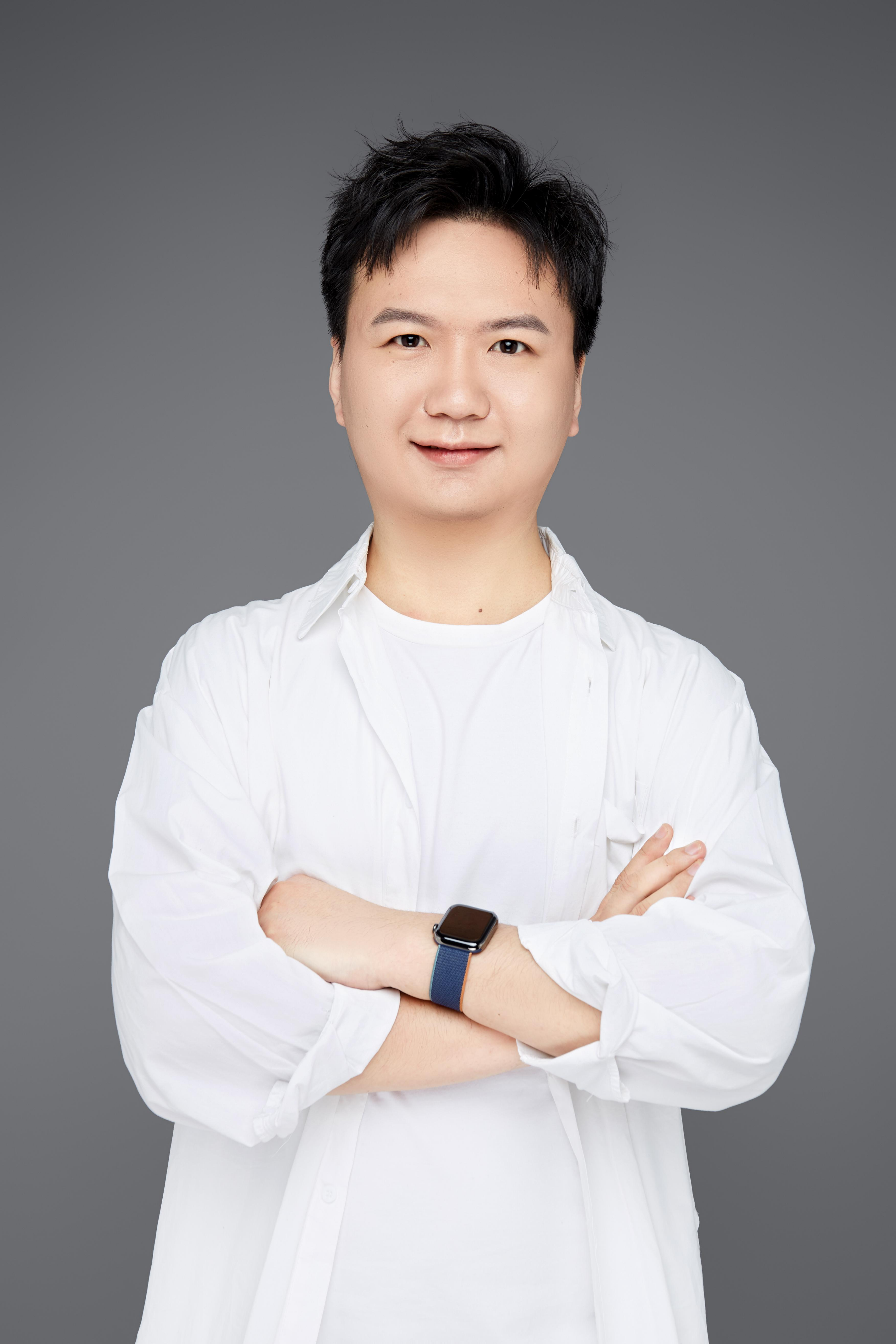}}]{Yuanhao Cui} (Member, IEEE) received the B.Eng. degree (Hons.) from Henan University and the Ph.D. degree from the Beijing University of Posts and Telecommunications (BUPT). He has been the CTO and the Co-Founder of two startup companies, where he has invested more than \$3 million dollars. He holds more than 20 granted patents. His research interests include precoding and protocol designs for ISAC. He is a member of the IMT-2030 (6G) ISAC Task Group. He received the Best Paper Award from IWCMC 2021. He is the Founding Secretary of the IEEE ComSoc ISAC Emerging Technology Initiative (ISAC-ETI) and the CCF Science Communication Working Committee. He was the Organizer and the Co-Chair for a number of workshops and special sessions in flagship IEEE conferences, including Mobicom, ICC, ICASSP, WCNC, and VTC. He is the Lead Guest Editor for the Special Issue on Integrated Sensing and Communication for 6G of the IEEE OPEN JOURNAL OF THE COMMUNICATIONS SOCIETY and the Guest Editor for the Special Issue on Integrated Sensing and Communications for Future Green Networks of the IEEE TRANSACTIONS ON GREEN COMMUNICATIONS AND NETWORKING.
\end{IEEEbiography}

\begin{IEEEbiography}[{\includegraphics[width=1in,height=1.25in,clip,keepaspectratio]{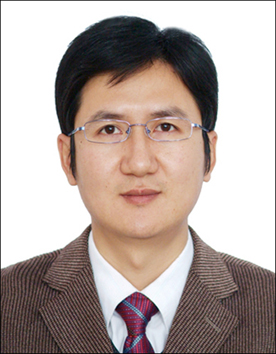}}]{Wei Wang} (Senior Member, IEEE) received PhD in Navigation, Guidance and Control from Harbin Engineering University (HEU), Heilongjiang, China, 2005. Postdoctoral research associate at Harbin Institute of Technology, China (July 2006 to April 2009). Associate Professor at Harbin Engineering University (August 2008 -- August 2010) and Academic Visitor at Loughborough University, UK (January 2010 -- December 2010). Professor at Harbin Engineering University (September, 2011 to now). 
	
He has published about more than 80 referred journal and conference papers. He is a Senior Member of the Institution of Electrical and Electronic Engineers (IEEE). His current research interests include Signal Processing for Wireless Navigation Systems and MIMO Radar.
\end{IEEEbiography}

\begin{IEEEbiography}[{\includegraphics[width=1in,height=1.25in,clip,keepaspectratio]{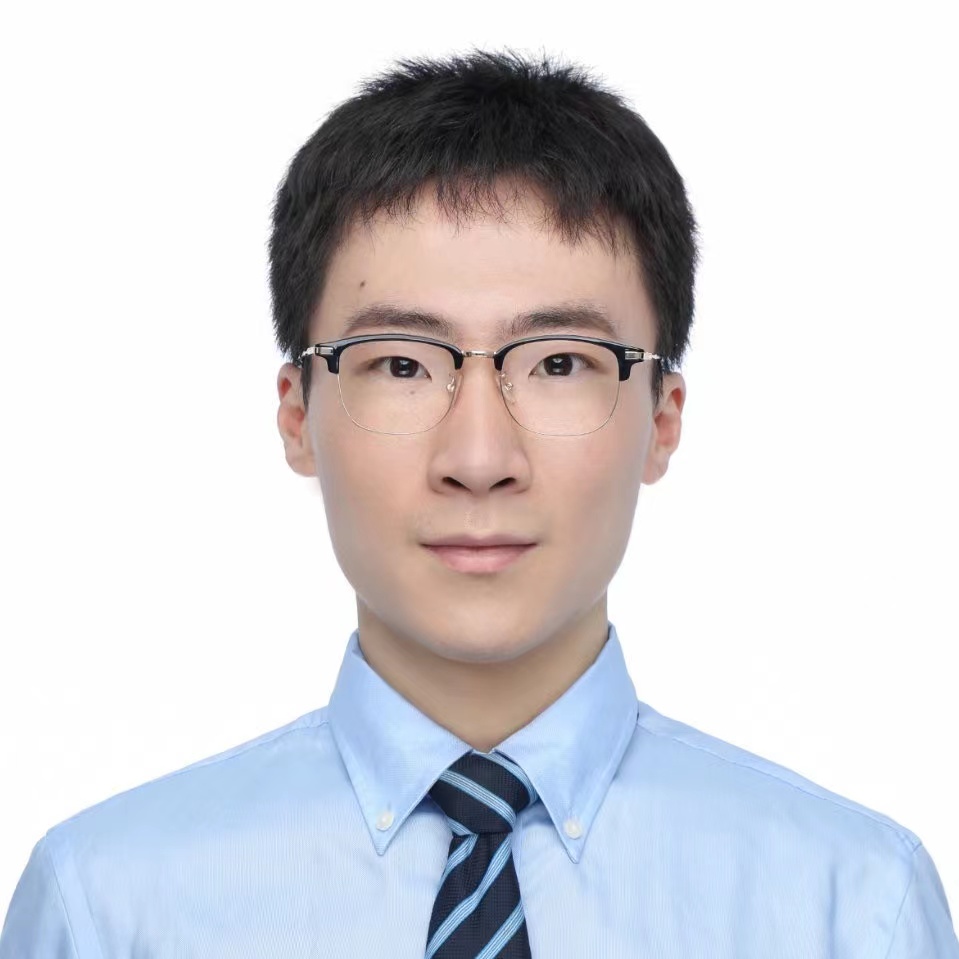}}]{Kaifeng Han} (Member, IEEE) received his Ph.D. degree from the University of Hong Kong in 2019, and the B.Eng. (first-class hons.) from the Beijing University of Posts and Telecommunications (BUPT) and Queen Mary University of London (QMUL) in 2015, all in electrical engineering. He has been an senior engineer at the China Academy of Information and Communications Technology (CAICT). His research interests focus on integrated sensing and communications, wireless AI for 6G.
\end{IEEEbiography}

\begin{IEEEbiography}[{\includegraphics[width=1in,height=1.25in,clip,keepaspectratio]{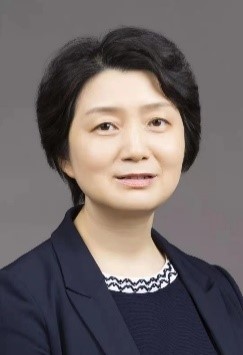}}]{Zhiqin Wang} is the Vice President of the China Academy of Information and Communications Technology (CAICT), the Professor-level senior engineer. She is currently the Chair of the Wireless Technical Committee of China Communications Standards Association (CCSA) and the Director of the Wireless and Mobile Technical Committee of China Institute of Communications (CIC). She is also serving as the Chair of the IMT-2020 (5G) Promotion Group and Chair of the  IMT-2030(6G) Promotion Group in China. She has contributed to the design, standardization, and development of 3G (TD-SCDMA), 4G (TD-LTE), and 5G mobile communication systems. She has authored or co-authored more than 60 research papers/articles, and over 20 patents in this area. Her achievements have received multiple top awards and honors by China central government, including the Grand Prize of the National Award for Scientific and Technological Progress in 2016 (the highest Prize in China), the First Prize of the National Award for Scientific and Technological Progress once, the Second Prize of the National Award for Scientific and Technological Progress twice, the National Innovation Award (once), and the Ministerial Science and Technology Awards many times. Her current research interests include the theoretical research, standardization, and industry development for 5G-Advanced and 6G.
\end{IEEEbiography}

%

%

\end{document}